\documentclass[twocolumn]{aastex63}

\usepackage{natbib}
\usepackage{color}
\usepackage{mathtools}
\usepackage{hyperref} 

\def    \apjl  		{\rm {ApJL}}

\def    \apjl  		{\rm {ApJL}}

\def	\cm		{\,{\rm {cm}}}
\def	\K		{\,{\rm {K}}}
\def	\g		{\,{\rm {g}}}
\def	\mum	{\,{\mu \rm{m}}}

\def \bea {\begin{eqnarray}}
\def \ena {\end{eqnarray}}

\def	\AAt	 {\,{\rm \AA}}		

\def	\B	{{\rm B}}

\def	\C	{{\rm C}}

\def	\cm	{\,{\rm cm}}
\def	\d	{{\rm d}}

\def	\D	{{\rm D}}

\def	\erg	{\,{\rm erg}}

\def	\g	{\,{\rm g}}
\def	\gas	{\,{\rm gas}}
\def	\km	{\,{\rm km}}
\def	\G	{{\rm G}}

\def	\H	{{\rm H}}

\def	\IR	{{\rm IR}}

\def	\N	{{\rm N}}

\def	\s	{\,{\rm s}}

\def	\AU	{\,{\rm AU}}

\def	\V	{{\rm V}}

\def	\rad	{{\rm rad}}
\def	\yr	    {\,{\rm yr}}

\def    \gas     	{{\rm gas}}

\begin{document}
\shorttitle{Rotational Desorption of Ice Mantles}
\shortauthors{Hoang and Tram}
\title{Rotational Desorption of Ice Mantles from Suprathermally Rotating Grains around Young Stellar Objects}
\author{Thiem Hoang}
\affil{Korea Astronomy and Space Science Institute, Daejeon 34055, Republic of Korea; \href{mailto:thiemhoang@kasi.re.kr}{thiemhoang@kasi.re.kr}}
\affil{University of Science and Technology, Korea, (UST), 217 Gajeong-ro Yuseong-gu, Daejeon 34113, Republic of Korea}

\author{Le Ngoc Tram}
\affil{SOFIA-USRA, NASA Ames Research Center, MS 232-11, Moffett Field, CA 94035, USA}
\affil{University of Science and Technology of Hanoi, VAST, 18 Hoang Quoc Viet, Hanoi, Vietnam}

\begin{abstract}
Ice mantles on dust grains play a central role in astrochemistry. Water and complex organic molecules (COMs) are thought to first form on the ice mantles and subsequently are released into the gas phase due to star formation activity. However, the critical question is whether ice mantles can survive stellar radiation when grains are being heated from $T_{d}\sim 10\K$ to $\gtrsim 100$ K. In this paper, we first study the effect of suprathermal grain rotation driven by the intense radiation of young stellar objects (YSOs) on the ice mantles. We find that the entire ice mantles can be disrupted into small fragments by centrifugal stress before the water ice and COMs desorb via thermal sublimation. We then study the consequence of resulting ice fragments and find that tiny fragments of radius $a \lesssim 10$\AAt~ exhibit transient release of COMs due to thermal spikes, whereas larger fragments can facilitate thermal sublimation at much higher rates than from the original icy grain or the same rate but with temperatures of $\sim 20-40$ K lower. We find that rotational desorption is efficient for hot cores/corinos from the inner to outer regions where the temperature drops to $T_{\gas}\sim 40\K$ and $n_{\H}\sim 10^{4}\cm^{-3}$. We discuss the implications of this mechanism for desorption of COMs and water ice in various environments, including outflow cavity walls, photodissociation regions, and protoplanetary disks. Finally, we show that very large aggregate grains can be disrupted into individual icy grains via rotational disruption mechanism, followed by rotational desorption of ice mantles.
 \end{abstract}

\section{Introduction}\label{sec:intro}
Water molecules are essential for life, and complex organic molecules (COMs referred to organic molecules containing $\ge 6$ atoms, such as CH$_3$OH, CH$_3$OCH$_3$, HCOOCH$_3$, and C$_2$H$_5$OH, CH$_3$CH$_2$CN), are the building blocks of life. Understanding where and how such molecules are formed and released into the gas phase is a key question in astrochemistry. COMs are increasingly observed in the environs of young stellar objects (YSOs), including hot cores/corinos around high-mass/low-mass protostars and protoplanetary disks (see \citealt{Herbst:2009go} and \citealt{vanDishoeck:2014cu} for recent reviews). In the formation process of water and COMs, the ice mantle of dust grains is known to play a central role (see \citealt{vanDishoeck:2014cu} and \citealt{2018IAUS..332....3V} for recent reviews).

The popular scenario to form COMs in hot cores/corinos involves three phases, including cold, warm, and hot phases \citep{Herbst:2009go}. During the initial cold phase, COMs may first be formed in the cold molecular core (\citealt{2016ApJ...830L...6J}) but are frozen in the icy grain mantle during the cloud collapse process (zeroth-generation species). During the warm phase, i.e., after star formation, the ice mantle is warmed up from $\sim 10$ K to $\sim 100$ K by protostellar radiation, which increases the mobility of simple molecules frozen in the ice mantle and finally form COMs (first-generation species; see, e.g., \citealt{2008ApJ...682..283G}). During the hot phase where icy grain mantles are heated to $100 - 300$ K, thermal sublimation of ice mantles (\citealt{1987ApJ...315..621B}; \citealt{1988MNRAS.231..409B}; \citealt{Bisschop:2007cu}) can release molecules (CH$_3$OH, NH$_3$), which trigger gas-phase chemistry at high temperatures and form in-situ COMs (second-generation species; see \citealt{1992ApJ...399L..71C}). 

Given the crucial importance of the icy grain mantles on the formation and desorption of water and COMs, the remaining question is whether the ice mantles can still survive in the intense radiation field of YSOs during the warm phase when thermal sublimation is not yet effective.

Previous research on thermal and non-thermal desorption of molecules from the grain mantle assumed that grains are at rest, which is contrary to the fact that grains are rapidly rotating due to collisions with gas atoms and interstellar photons (\citealt{1998ApJ...508..157D}; \citealt{Hoang:2010jy}). To understand how molecules form on the grain surfaces and are returned to the gas, the effect of grain rotation of gas-grain chemistry must be quantified. The goal of this paper is to quantify the effect of grain rotation on the desorption of ice mantles from the grain surface.

Interstellar dust grains are widely known to be rotating suprathermally, as required to reproduce starlight polarization and far-IR/submm polarized dust emission (see \citealt{Andersson:2015bq} and \citealt{LAH15} for reviews). Indeed, \cite{1979ApJ...231..404P} first suggested that dust grains can be spun-up to suprathermal rotation (with velocities larger than grain thermal velocity) by various mechanisms, including the formation of hydrogen molecules on the grain surface. 

In particular, modern astrophysics establishes that dust grains of irregular shapes can rotate suprathermally due to radiative torques (RATs) arising from their interaction with an anisotropic radiation field (\citealt{1996ApJ...470..551D}; \citealt{2007MNRAS.378..910L}; \citealt{Hoang:2008gb}; \citealt{2009ApJ...695.1457H}; \citealt{Herranen:2019kj}) or mechanical torques induced by an anisotropic gas flow (\citealt{2007ApJ...669L..77L}; \citealt{2018ApJ...852..129H}). In an intense radiation field, \cite{Hoang:2019da} discovered that irregular grains could be spun-up to extremely fast rotation such that the centrifugal stress can exceed the tensile strength of the grain material, breaking the original grain into many fragments. Therefore, in star-forming regions and photodissociation regions, we expect that the radiation intensity is sufficiently intense such that the centrifugal force can disrupt the ice mantle from the grain core into small icy fragments. Consequently, sublimation from such tiny fragments would occur at a much higher rate than from the original large grain. This issue will be quantified in the present paper. 

The structure of the paper is as follows. In Section \ref{sec:desorp_grain} we will introduce the rotational desorption mechanism of ice mantles from suprathermally rotating dust grains spun-up by RATs due to centrifugal stress. We will demonstrate that resulting small fragments can induce faster evaporation of COMs than classical thermal sublimation in Section \ref{sec:results}. In Section \ref{sec:hotcore}, we apply rotational desorption mechanism for hot cores/corinos around protostars. Discussion for other environments, including photodissociation regions (PDRs) and protoplanetary disks (PPDs), and the effect of grain evolution for rotational desorption are presented in Section \ref{sec:discuss}. Major findings are summarized in Section \ref{sec:summ}.

\section{Rotational desorption mechanism of icy grain mantles}\label{sec:desorp_grain}

\subsection{Ice mantles on grain surface}
The formation of an ice mantle due to accretion of gas molecules on the grain surface is expected to occur in cold and dense regions of hydrogen density $n_{\H}=n(\H)+2n(\H_{2}) \sim 10^{3}-10^{5}\cm^{-3}$ or the visual extinction $A_{V}> 3$ (\citealt{1983Natur.303..218W}). Ice mantles on the grain surface are expected to have layer structures because water and CO condense at different temperatures, where the first layer is dominated by water ice and the second layer is dominated by CO ice (see \citealt{2010ApJ...716..825O}). The detection of strongly polarized H$_{2}$O and CO ice absorption features (\citealt{1996ApJ...465L..61C}; \citealt{2008ApJ...674..304W}) demonstrates that icy grain mantles have non-spherical shape and are aligned with magnetic fields (see \citealt{LAH15} for a review).

The mantle thickness can vary with the size of dust grains due to their difference in grain temperature (see \citealt{2016ApJ...817..146P}). In dense clouds, one expects the ice mantle to have $\sim$ 100 monolayers on large grains ($a\gtrsim 0.1\mum$) and about 30-50 monolayers on small grains, assuming gas density $n_{\H}\sim 10^{4}\cm^{-3}$. At high densities of $\sim 10^{7}\cm^{-3}$, the ice mantle can grow to $\sim$ 380 monolayers after $\sim$ 1 Myr \citep{2011ApJ...735...15G}.

\begin{figure}
\centering
\includegraphics[scale=0.3]{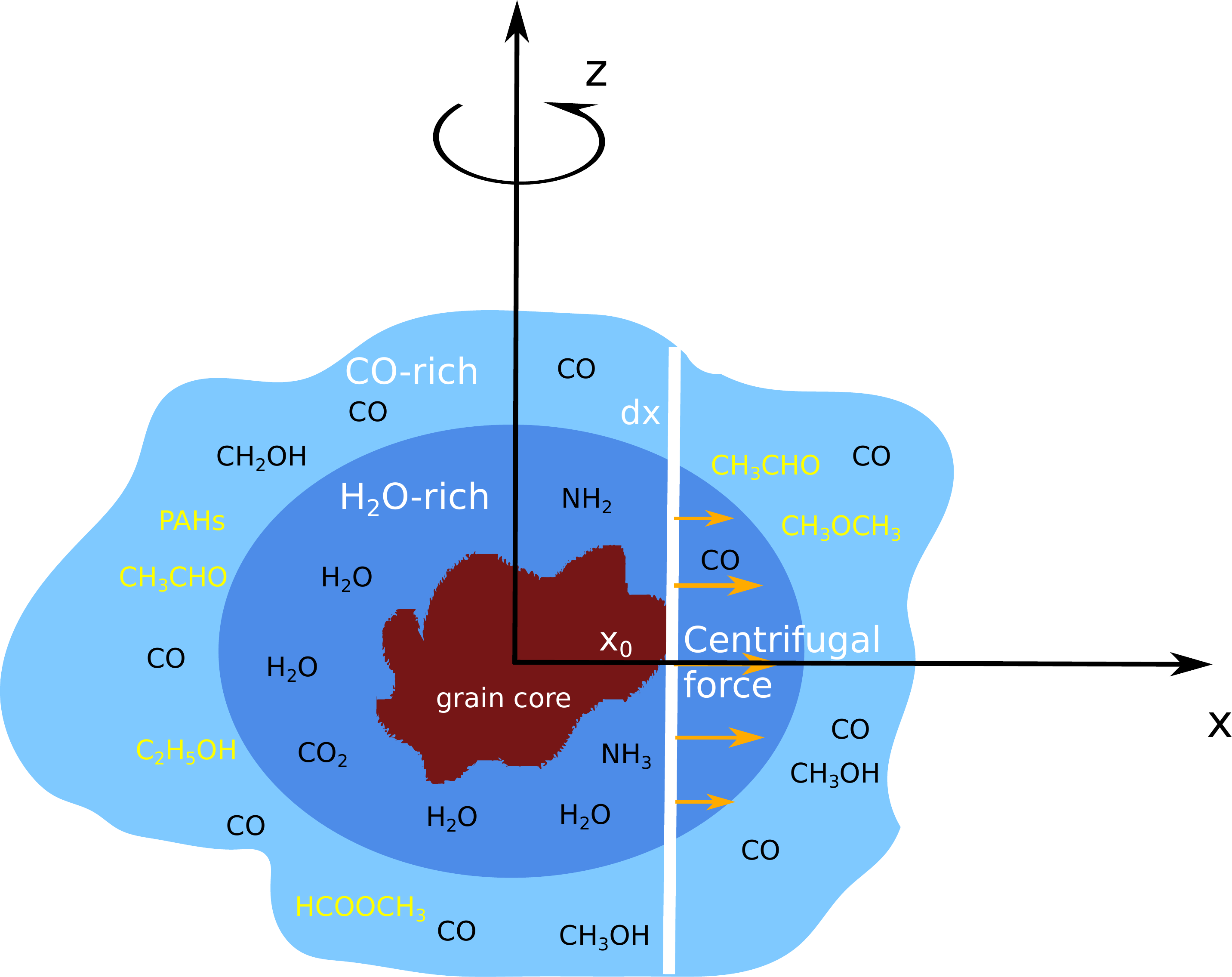}
\caption{A schematic illustration of a rapidly spinning core-mantle grain of irregular shape, comprising an icy water-rich (blue) and CO-rich (orange) mantle layers. The core is assumed to be compact silicate material, and complex organic molecules are formed in the ice mantle of the core. Centrifugal force field on a slab $dx$ is illustrated, which acts to pull off the ice mantle from the grain core at sufficiently fast rotation.}
\label{fig:grain_mod}
\end{figure}

Here we consider a grain model consisting of an amorphous silicate core covered by a double-layer ice mantle (see Figure \ref{fig:grain_mod}). Let $a_{c}$ be the radius of silicate core and $\Delta a_{m}$ be the average thickness of the mantle. The exact shape of icy grains is unknown, but we can assume that they have irregular shapes as required by polarization data \citep{1996ApJ...465L..61C}. Thus, one can define an effective radius of the grain, $a$, which is defined as the radius of the sphere with the same volume as the grain. The effective grain size is $a \approx a_{c}+\Delta a_{m}$. The silicate and carbonaceous core perhaps are assumed to have a typical radius of $0.05 \mum$ \citep{1989IAUS..135..345G}. In the following, we first assume the maximum size of core-mantle grains is $a_{\max}\sim 1\mum$ and postpone discussion of grain growth to Section \ref{sec:discuss}).

\subsection{Suprathermal rotation of dust grains by RATs}
We consider an anisotropic radiation field of the anisotropy degree $\gamma$. Let $u_{\lambda}$ be the spectral energy density of radiation field at wavelength $\lambda$. The energy density of the radiation field is then $u_{\rad}=\int u_{\lambda}d\lambda$. To describe the strength of a radiation field, let define $U=u_{\rm rad}/u_{\rm ISRF}$ with 
$u_{\rm ISRF}=8.64\times 10^{-13}\erg\cm^{-3}$ being the energy density of the average interstellar radiation field (ISRF) in the solar neighborhood as given by \cite{1983A&A...128..212M}. Thus, the typical value for the ISRF is $U=1$. Let $\overline{\lambda}=\int \lambda u_{\lambda}d\lambda/u_{\rm rad}$ be the mean wavelength of the radiation field. For the ISRF, $\bar{\lambda}\sim 1.2\mum$ (\citealt{1997ApJ...480..633D}).

As shown in \cite{Hoang:2019da}, for the radiation source of constant luminosity, i.e., constant radiative torque $\Gamma_{\rm RAT}$ (\citealt{1996ApJ...470..551D}; \citealt{2007MNRAS.378..910L}), the grain angular velocity is steadily increased over time as
\bea
\omega(t)=\omega_{\rm RAT}\left[1-\exp\left(-\frac{t}{\tau_{\rm damp}}\right)\right],
\ena
where
\bea
\omega_{\rm RAT}=\frac{\Gamma_{\rm RAT}\tau_{\rm damp}}{I}\label{eq:omega_RAT0}
\ena
is the terminal angular velocity at time much larger than damping time, $t\gg \tau_{\rm damp}$, which is considered the maximum rotational rate spun-up by RATs (\citealt{Hoang:2019da}). 

Above, $\tau_{\rm damp}=t_{\gas}^{-1}+\tau_{\IR}^{-1}=\tau_{\gas}^{-1}(1+F_{\rm IR})$ is the total damping rate where $t_{\gas}$ is the rotational damping time due to gas collisions, and $F_{\IR}$ is the grain rotational damping coefficient due to infrared emission (see \citealt{1998ApJ...508..157D}). For a gas with He of $10\%$ abundance, the characteristic damping time due to collisions is
\bea
\tau_{\gas}&=&\frac{3}{4\sqrt{\pi}}\frac{I}{1.2n_{\rm H}m_{\rm H}
v_{\rm th}a^{4}}\nonumber\\
&\simeq& 8.74\times 10^{4}a_{-5}\hat{\rho}\left(\frac{30\cm^{-3}}{n_{\H}}\right)\left(\frac{100\K}{T_{\gas}}\right)^{1/2}~{\rm yr},~~
\ena
where $a_{-5}=a/(10^{-5}\cm)$, $\hat{\rho}=\rho/(3\g\cm^{-3})$ with $\rho$ being the dust mass density, $v_{\rm th}=\left(2k_{\B}T_{\rm gas}/m_{\rm H}\right)^{1/2}$ is the thermal velocity of gas atoms of mass $m_{\rm H}$ in a plasma with temperature $T_{\gas}$ and density $n_{\H}$, and spherical grains are assumed (\citealt{2009ApJ...695.1457H}; \citealt{1996ApJ...470..551D}).

For grains in thermal equilibrium due to starlight heating and radiative cooling, one obtains (see \citealt{1998ApJ...508..157D}):
\bea
F_{\rm IR}\simeq \left(\frac{0.4U^{2/3}}{a_{-5}}\right)
\left(\frac{30 \cm^{-3}}{n_{\H}}\right)\left(\frac{100 \K}{T_{\gas}}\right)^{1/2}.\label{eq:FIR}
\ena 

Following \cite{2019ApJ...876...13H}, the maximum rotation rate of grains spun-up by RATs is given by
\bea
\omega_{\rm RAT}&\simeq &9.6\times 10^{8}\gamma a_{-5}^{0.7}\bar{\lambda}_{0.5}^{-1.7}\nonumber\\
&\times&\left(\frac{U}{n_{1}T_{2}^{1/2}}\right)\left(\frac{1}{1+F_{\rm IR}}\right)\rad\s^{-1},~~~\label{eq:omega_RAT1}
\ena
for grains with $a\le \bar{\lambda}/1.8$, and
\bea
\omega_{\rm RAT}&\simeq &1.78\times 10^{10}\gamma a_{-5}^{-2}\bar{\lambda}_{0.5}\nonumber\\
&&\times\left(\frac{U}{n_{1}T_{2}^{1/2}}\right)\left(\frac{1}{1+F_{\rm IR}}\right)\rad\s^{-1},~~~\label{eq:omegaRAT2}
\ena
for grains with $a> \overline{\lambda}/1.8$. Above, $n_{1}=n_{\H}/(10\cm^{-3}), T_{2}=T_{\gas}/(100\K)$, and $\bar{\lambda}_{0.5}=\bar{\lambda}/(0.5\mum)$. The rotation rate depends on the parameter $U/n_{\H}T_{\gas}^{1/2}$ and $F_{\rm IR}$. For $U\gg1$, $\omega_{\rm RAT}$ is much larger than the thermal angular velocity of grains $\omega_{T}=(2kT_{\gas}/I)^{1/2}\sim 2\times 10^{5}a_{-5}^{-5/2}T_{2}^{1/2}\rad\s^{-1}$, which is referred to as suprathermal rotation.

For convenience, let $a_{\rm trans}=\bar{\lambda}/1.8$ which denotes the grain size at which the RAT efficiency changes between the power law and flat stages (see e.g., \citealt{2007MNRAS.378..910L}; \citealt{Hoang:2019da}), and $\omega_{\rm RAT}$ changes from Equation (\ref{eq:omega_RAT1}) to (\ref{eq:omegaRAT2}).

We note that the RAT efficiency weakly depends on the composition of dust grains as shown in \cite{2007MNRAS.378..910L} and \cite{Herranen:2019kj}. Thus, Equations (\ref{eq:omega_RAT1}) and (\ref{eq:omegaRAT2}) can be applicable for both silicate and carbonaceous grains.

\subsection{Rotational desorption of ice mantles}
\subsubsection{Centrifugal stress and tensile strength of ice-mantles}

Assuming that the grain is rotating around the axis of maximum inertia moment, denoted by z-axis, with angular velocity $\omega$. This assumption is valid for suprathermal rotating grains in which internal relaxation can rapidly induce the perfect alignment of the axis of the major inertia with the angular momentum which corresponds to the minimum rotational energy state \citep{1979ApJ...231..404P}. Let us consider a slab $dx$ at distance $x$ from the center of mass. The average tensile stress due to centrifugal force $dF_{c}$ acting on a plane located at distance $x_{0}$ is equal to
\bea
dS=\frac{\omega^{2}x dm}{\pi (a^{2}-x_{0}^{2})}= \frac{\rho_{\rm ice} \omega^{2}(a^{2}-x^{2})xdx}{a^{2}-x_{0}^{2}},
\ena
where the mass of the slab $dm=\rho_{\rm ice} dAdx$ with $dA=\pi(a^{2}-x^{2})$ the area of the circular slab, $\rho_{\rm ice}$ is the mass density of the ice mantle, which is $\rho_{\rm ice}\sim 1\g\cm^{-3}$ for pure ice.  

The surface average tensile stress is then given by
\bea
S_{x}&=&\int_{x_{0}}^{a} dS_{x}=\frac{\rho_{\rm ice}\omega^{2}a^{2}}{2}\int_{x_{0}/a}^{1}\frac{(1-u)du}{1-u_{0}}\nonumber\\
&=& \frac{\rho_{\rm ice} \omega^{2}a^{2}}{4}\left(\frac{(1-u_{0})^{2}}{1-u_{0}}\right)=\frac{\rho_{\rm ice} \omega^{2}a^{2}}{4}\left[1-\left(\frac{x_{0}}{a}\right)^{2}\right],~~~\label{eq:Smax}
\ena
where $u=x^{2}/a^{2}$. 

Equation (\ref{eq:Smax}) reveals that the tensile stress is maximum at the grain center and decreases with decreasing the mantle thickness $(a-x_{0})$. 

By plugging the numerical numbers into Equation (\ref{eq:Smax}), one obtains
\bea
S_{x}\simeq 2.5\times 10^{9}\hat{\rho}_{\rm ice}\omega_{10}^{2}a_{-5}^{2} \left[1-\left(\frac{x_{0}}{a}\right)^{2} \right] \erg \cm^{-3},\label{eq:Sx}
\ena
where $\hat{\rho}_{\rm ice}=\rho_{\rm ice}/(1\g\cm^{-3})$ and $\omega_{10}=\omega/(10^{10}\rm rad\s^{-1})$

The tensile strength of the bulk ice is $S\sim 2\times 10^{7}\erg\cm^{-3}$ at low temperatures. As the temperature increases to $200-300$ K, the tensile strength is reduced significantly to $5\times 10^{6}\erg\cm^{-3}$ \citep{Litwin:2012ii}. The adhesive strength between the ice mantle and the solid surface has a wide range, depending on the surface properties (\citealt{Itagaki:1983ud}; \citealt{Work:2018bu}). Here, we adopt a conservative value of $S_{\max}=10^{7}\erg\cm^{-3}$ for ice mantles for our numerical calculations. For the grain core, a higher value of $S_{\max}=10^{9}\erg\cm^{-3}$ is adopted.

When the rotation rate is sufficiently high such as the tensile stress exceeds the maximum limit of the ice mantle, $S_{\rm max}$, the grain is disrupted. The critical rotational velocity is determined by $S_{x}=S_{\rm max}$:
\bea
\omega_{\rm disr}&=&\frac{2}{a(1-x_{0}^{2}/a^{2})^{1/2}}\left(\frac{S_{\max}}{\rho_{\rm ice}} \right)^{1/2}\nonumber\\
&\simeq& \frac{6.3\times 10^{8}}{a_{-5}(1-x_{0}^{2}/a^{2})^{1/2}}\hat{\rho}_{\rm ice}^{-1/2}S_{\max,7}^{1/2}~\rad\s^{-1},\label{eq:omega_disr}
\ena
where $S_{\max,7}=S_{\max}/(10^{7} \erg \cm^{-3})$.

Although the detail of rotational disruption of dust grains is not yet studied, the disruption process perhaps undergoes the following steps. When the grain rotation rate is increased to the critical disruption limit, the ice mantle near the equator is detached because the centrifugal stress is equal to the tensile strength of the mantle onto the grain core. As the rotation rate is further increased beyond $\omega_{\rm disr}$, the centrifugal stress exceeds the ice tensile strength that holds the different parts of the mantle together, resulting in the disruption of the mantle into small fragments. 

Above, we assume that the grain is spinning along the principal axis of maximum inertia moment. This assumption is valid because internal relaxation within the rapidly spinning grain due to Barnett effect rapidly brings the grain axis to be aligned with its angular momentum (\citealt{1979ApJ...231..404P}; \citealt{1999MNRAS.305..615R}).

\subsubsection{Desorption sizes of ice mantles}

The critical grain size of rotational desorption can be found by setting $\omega_{\rm RAT}=\omega_{\rm disr}$. From Equations (\ref{eq:omega_RAT1}-\ref{eq:omegaRAT2}) and (\ref{eq:omega_disr}), one can see that when the grain size increases, $\omega_{\rm RAT}$ rapidly increases and then intersects with $\omega_{\rm disr}$ at $a=a_{\rm disr}$, assuming a sufficiently strong radiation fields. When the grain size continues to increase beyond $a_{\rm trans}$, $\omega_{\rm RAT}$ then declines rapidly as $\propto a^{-2}$ and then intersect $\omega_{\rm disr}\propto a^{-1}$ again at some large grain size, producing a second intersection. The first intersection determines the critical size above which grains are disrupted, whereas the second intersection determines the maximum size that grains still can be disrupted.

\begin{figure*}
\centering
\includegraphics[scale=0.49]{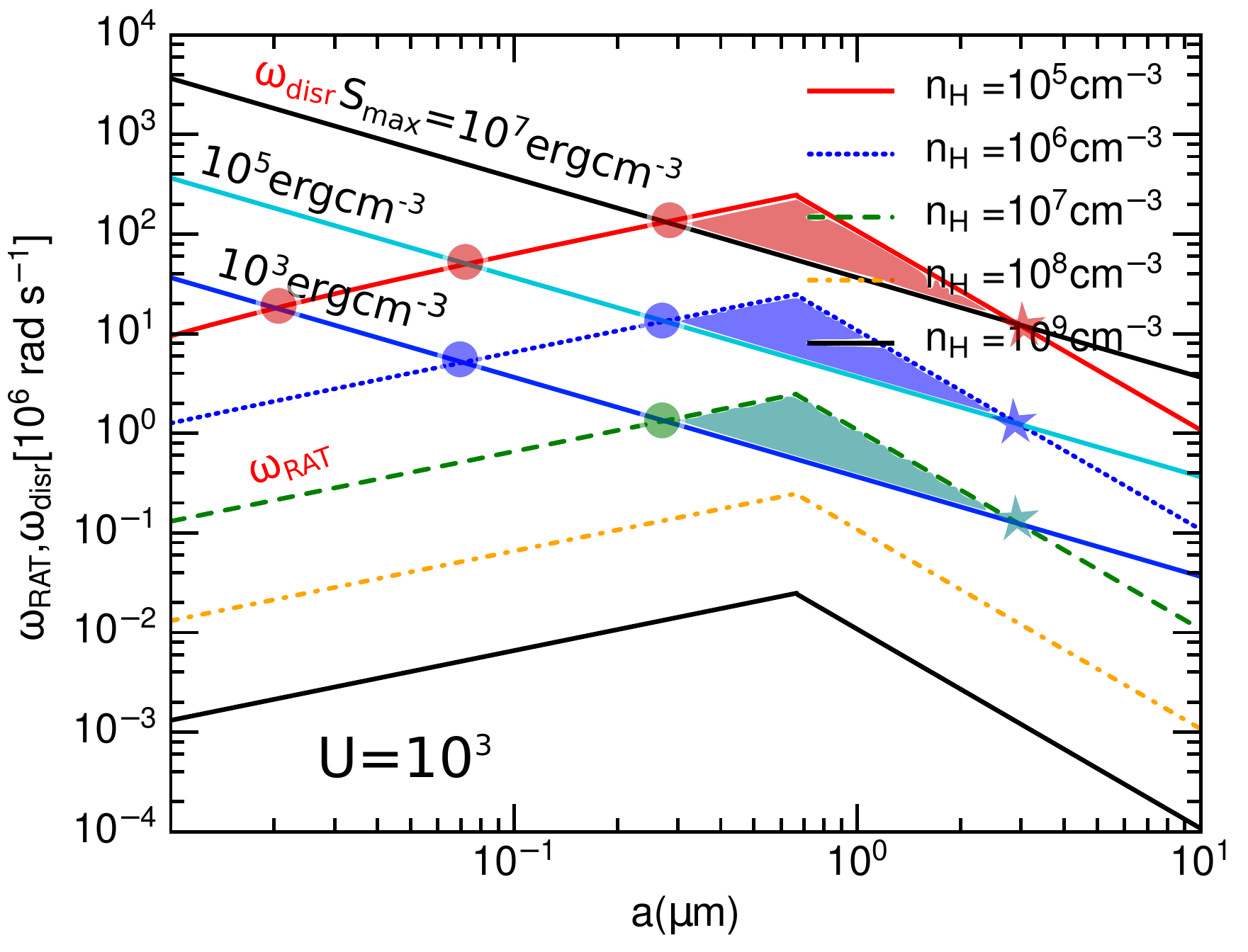}
\includegraphics[scale=0.49]{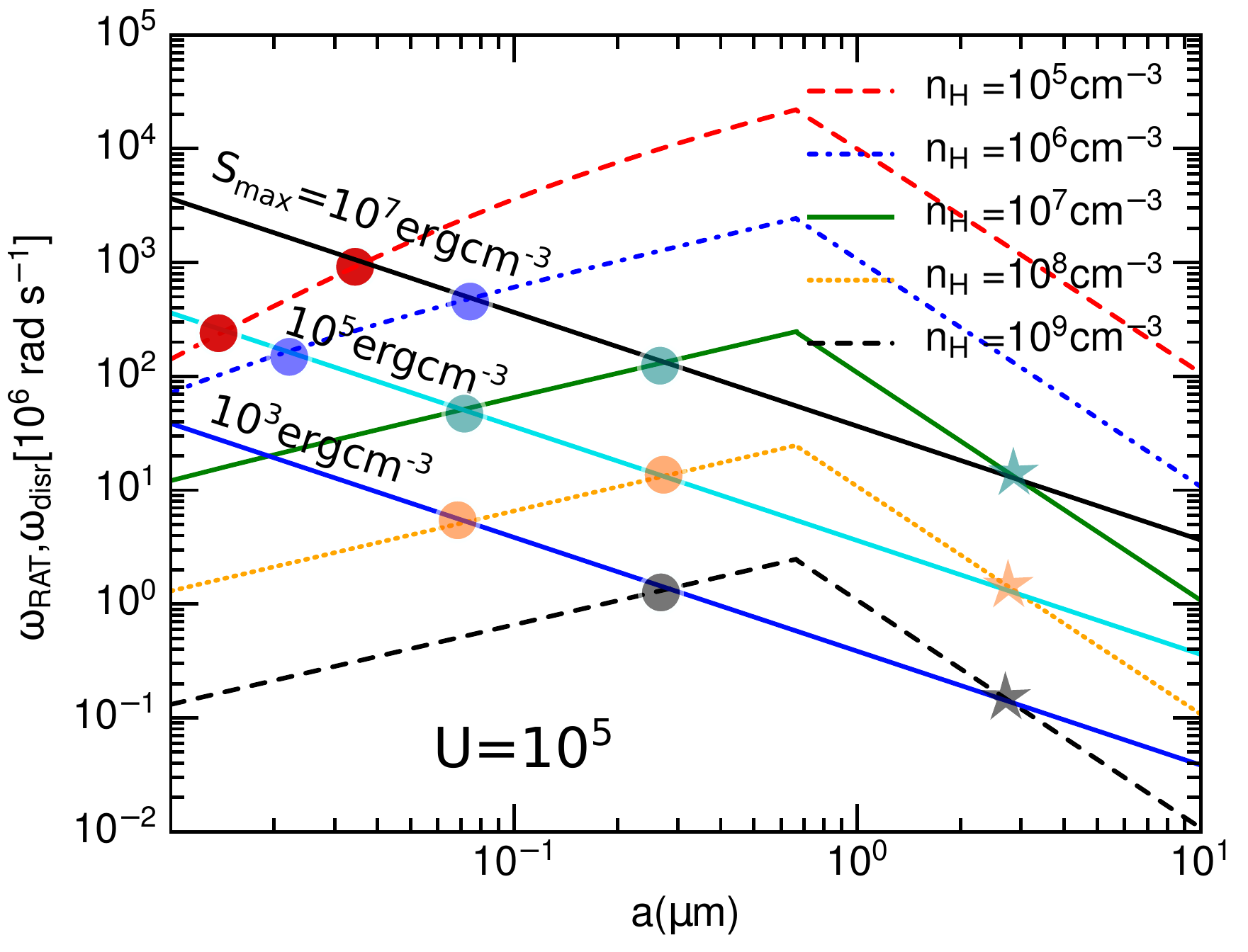}
\caption{The grain rotation rate spun-up by RATs ($\omega_{\rm RAT}$) and disruption rate ($\omega_{\rm disr}$) as functions of the grain size for different gas densities. Two values of the radiation strength $U=10^{3}$ (left panel) and $U=10^{5}$ (right panel) and three values of tensile strengths $S_{\max}=10^{3}, 10^{5}, 10^{7}\erg\cm^{-3}$ are considered. The peak of $\omega_{\rm RAT}$ occurs at $a=a_{\rm trans}$. The intersection of $\omega_{\rm RAT}$ and $\omega_{\rm disr}$ can occur at a lower grain size (marked by a circle) and an upper size (marked by a star), and the shaded area denotes the range of grain sizes in which grains are disrupted by RATs.}
\label{fig:omegaRAT_disr}
\end{figure*}

Figure \ref{fig:omegaRAT_disr} shows the rotation rate vs. disruption rate for different gas densities, tensile strengths, assuming the radiation strength of $U=10^{3}$ and $10^{5}$. The intermediate strength of $S_{\max}\sim 10^{5}\erg\cm^{-3}$ is expected for composite grains (\citealt{2019ApJ...876...13H}; see \citealt{Gundlach:2018cu} for experimental results). The intersection between $\omega_{\rm RAT}$ and $\omega_{\rm disr}$ occurs at a lower grain size ($a_{\rm disr}$) and a very large grain size ($a_{\rm disr,max}$). The range $[a_{\rm disr}-a_{\rm disr,max}]$ describe the sizes of grains that are rotationally disrupted, which is referred to as the disruption size range. When the disruption size is smaller than $a_{\rm trans}$, the disruption size range becomes broader. On the other hand, when the disruption size $a_{\rm disr}\rightarrow a_{\rm trans}$, the disruption size range is reduced to $a=a_{\rm trans}$.

The grain size at the first intersection can be obtained from Equation (\ref{eq:omega_RAT1}) and (\ref{eq:omega_disr}), which is given by
\bea
a_{\rm disr}&\simeq&0.13\gamma^{-1/1.7}\bar{\lambda}_
{0.5}(S_{\max,7}/\hat{\rho}_{\rm ice})^{1/3.4}\nonumber\\
&&\times (1+F_{\rm IR})^{1/1.7}\left(\frac{n_{1}T_{2}^{1/2}}{U}\right)^{1/1.7}\mum,~~~\label{eq:adisr_low}
\ena
for $a_{\rm disr}\lesssim a_{\rm trans}$ and $x_{0}\ll a$, which depends on the local gas density and temperature due to gas damping. The equation indicates that all grains in the size range $a_{\rm trans}>a>a_{\rm disr}$ would be disrupted. 

To determine the maximum size of grains that can still be disrupted by centrifugal stress, one needs to compare $\omega_{\rm RAT}$ from Equation (\ref{eq:omegaRAT2}) with $\omega_{\rm disr}$, which yields:
\bea
a_{\rm disr,max}\simeq2.9\gamma\bar{\lambda}_{0.5}\left(\frac{U}{n_{1}T_{2}^{1/2}}\right)\left(\frac{1}{1+F_{\rm IR}}\right)\hat{\rho}_{\rm ice}^{1/2}S_{\max,7}^{-1/2}\mum.~~~~~\label{eq:adisr_up}
\ena 

For the standard parameters of the diffuse interstellar medium (ISM), one gets $a_{\rm disr,max}\sim 2.9\mum$ for the typical physical parameters in Equation (\ref{eq:adisr_up}). This is much larger than the maximum grain size of $a_{\max}\sim 0.25-0.3\mum$ obtained from modeling of observational data (\citealt{Mathis:1977p3072}; \citealt{1995ApJ...444..293K}; \citealt{2009ApJ...696....1D}). So, all available grains of $a\gtrsim a_{\rm disr}$ are disrupted. In dense regions, grains are expected to grow to large sizes due to coagulation and accretion (e.g., \citealt{Chokshi:1993p3420}; \citealt{Ossenkopf:1993p4016}). Therefore, not all grains of $a\gtrsim a_{\rm disr}$ can be disrupted, and we will find both $a_{\rm disr}$ and $a_{\rm disr,max}$ for grains in star-forming regions. Due to dependence of $F_{\rm IR}$ on the grain size and $S_{x}$ on $x_{0}$, one must find $a_{\rm disr}$ and $a_{\rm disr,max}$ using numerical calculations instead of Equations (\ref{eq:adisr_low}) and (\ref{eq:adisr_up}).

Note that calculations of RATs for irregular shapes are limited to grains of size $a<\lambda/0.1$ (\citealt{2007MNRAS.378..910L}; \citealt{Herranen:2019kj}). Therefore, the actual maximum disruption size might be lower than given by Equation (\ref{eq:adisr_up}) if $a_{\rm disr, max}>\lambda/0.1$ because RATs of such very large grains are expected to decrease due to random effects when different photons scan the different facets of the grain.

\subsubsection{Desorption time of ice mantles}

In the absence of rotational damping, the characteristic timescale for rotational desorption can be estimated as:
\bea
t_{\rm disr,0}&=&\frac{I\omega_{\rm disr}}{dJ/dt}=\frac{I\omega_{\rm disr}}{\Gamma_{\rm RAT}}\nonumber\\
&\simeq& 0.6(\gamma U_{5})^{-1}\bar{\lambda}_{0.5}^{1.7}\hat{\rho}_{\rm ice}^{1/2}S_{\max,7}^{1/2}a_{-5}^{-0.7}{~\rm yr}\label{eq:tdisr}
\ena
for $a_{\rm disr}<a \lesssim a_{\rm trans}$, and
\bea
t_{\rm disr,0}\simeq& 0.04(\gamma U_{5})^{-1}\bar{\lambda}_{0.5}^{-1}\hat{\rho}_{\rm ice}^{1/2}S_{\max,7}^{1/2}a_{-5}^{2}{~\rm yr}
\ena
for $a_{\rm trans}<a<a_{\rm disr,max}$ where $U_{5}=U/10^{5}$.

In the presence of rotational damping, the disruption timescale can be obtained by solving $\omega(t)=\omega_{\rm disr}$, which yields
\bea
t_{\rm disr}&=&-\tau_{\rm damp}\ln \left(1-\frac{\omega_{\rm disr}}{\omega_{\rm RAT}}\right)\nonumber\\
&=&-\tau_{\rm damp}\ln \left(1-\frac{t_{\rm disr,0}}{\tau_{\rm damp}}\right),\label{eq:tdisr_exact}
\ena
which is applicable for $a_{\rm disr,max}>a>a_{\rm disr}$. Note that $t_{\rm disr}\rightarrow \infty$ for $a=[a_{\rm disr}, a_{\rm disr,max}]$ because it takes $t\gg t_{\rm damp}$ to reach $\omega=\omega_{\rm RAT}$. One see that $t_{\rm disr}$ returns to $t_{\rm disr,0}$ when $t_{\rm disr,0}\ll \tau_{\rm damp}$ which is achieved in strong radiation fields.

\section{Rotational desorption of ice mantles and subsequent evaporation}\label{sec:results}

In this section, we will use the theory from the previous section to quantify the rotational desorption of ice mantles into tiny icy fragments. We then demonstrate that the subsequent evaporation of molecules from resulting icy fragments is much faster than the evaporation from the original icy mantle grain. 

\subsection{Rotational desorption size of ice mantles}
We consider a unidirectional radiation field (i.e., $\gamma=1$) in which dust grains are illuminated by a central YSO and consider two different mean wavelengths of the radiation field $\bar{\lambda}=0.5\mum$ and $1.2\mum$ (e.g., typical interstellar radiation). Specifically, we consider a range of gas density $n_{\H}\sim 10^{5}-10^{9}\cm^{-3}$ and radiation strength $U\sim 10^{3}-10^{8}$ around YSOs. Due to such dense conditions, the gas and dust grain temperatures (of large grains $a=0.1\mum$) are similar. The value of $a_{\rm disr,max}$ is set to $a_{\max}=1\mum$ in the absence of the intersection between $\omega_{\rm RAT}$ and $\omega_{\rm disr}$. 

We first consider a core-mantle grain model with a fixed core radius $a_{\rm c}=0.05\mum$, and the mantle thickness can vary. 

Table \ref{tab:adisr_mantle} shows the disruption size of the ice mantle from a core-mantle grain for selected physical parameters and the radiation field of $\bar{\lambda}=0.5\mum$, assuming a fixed core radius $a_{c}=0.05\mum$.

\begin{table}
\begin{center}
\caption{\rm Disruption size of ice mantles from the grain with fixed core radius $a_{c}=0.05\mum$}\label{tab:adisr_mantle}
\begin{tabular}{l l l l l l} \hline\hline\\
\multicolumn{2}{l}{Gas density} & 
\multicolumn{2}{c}{$a_{\rm disr}(\mu m)$}\\
n$_{\H}(\cm^{-3})$ & $U=10^{3}$ & $10^{4}$ & $10^{5}$ & $10^{6}$ & $10^{7}$\cr
& $T_{d}(\K)\approx 52$ & $76$ & $112$ & $164$ & $240$\cr

\hline\\
$10^{5}$ &  0.2597 & 0.0849 & 0.0525 & 0.0506 & 0.0249\cr
$10^{6}$ &  ND$^a$ & 0.2680 & 0.0921 & 0.0522 & 0.0303\cr
$10^{7}$ & ND & ND & ND & 0.0979 & 0.0528\cr
$10^{8}$ & ND & ND & ND & ND & 0.1092\cr
\hline
\multicolumn{5}{l}{$^a$~ND: No Disruption}\cr
\cr
\hline\hline
\end{tabular}
\end{center}
\end{table}

\begin{figure*}
\includegraphics[scale=0.5]{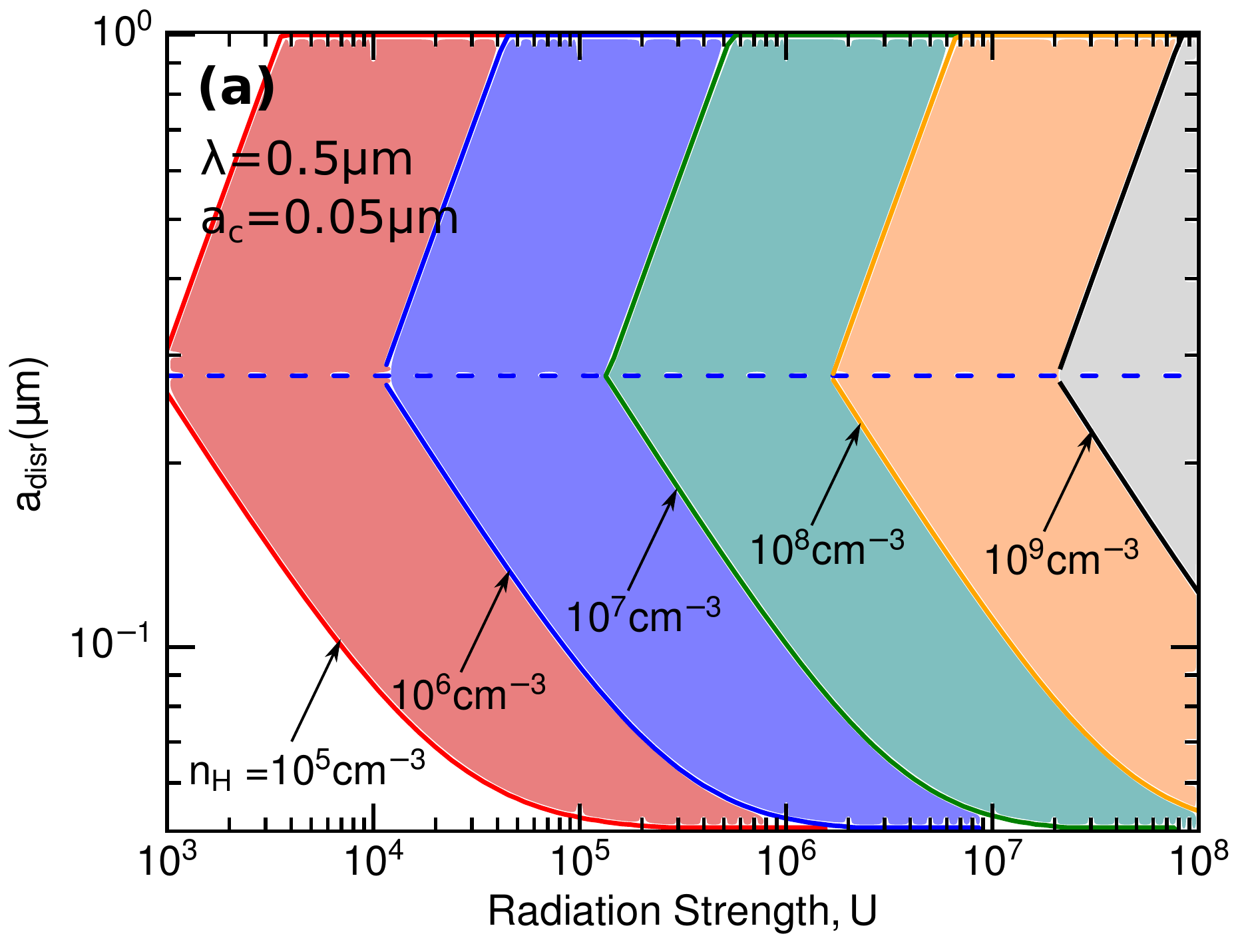}
\includegraphics[scale=0.5]{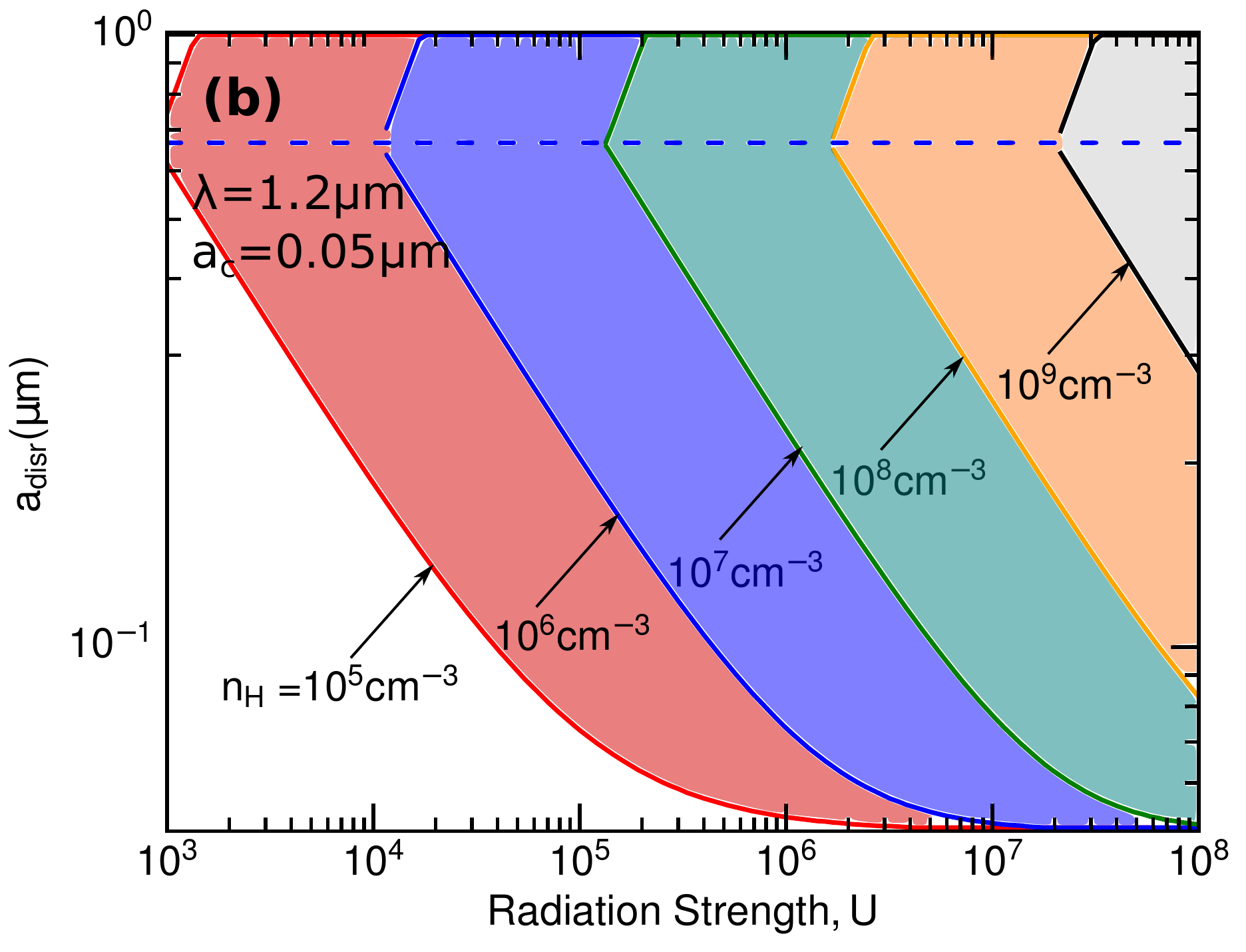}
\caption{Range of desorption sizes of ice mantles, constrained by $a_{\rm disr}$ (lower boundary) and $a_{\rm disr, max}$ (upper boundary), as a function of the grain temperature for the different gas densities for $\bar{\lambda}=0.5\mum$ (panel (a)) and $\bar{\lambda}=1.2\mum$ (panel (b)), assuming a fixed core radius $a_{c}$ and the varying mantle thickness. The horizontal dashed lines denote the transition size $a_{\rm trans}=\bar{\lambda}/1.8$. Shaded regions mark the range of grain sizes disrupted by RATD.}
\label{fig:adisr_U_coreshell}
\end{figure*}

Figure \ref{fig:adisr_U_coreshell} shows the range of grain disruption sizes ($a_{\rm disr},a_{\rm disr,max}$) as functions of the radiation strength $U$ for the different gas density, assuming $\bar{\lambda}=0.5\mum$ (panel (a)) and $\bar{\lambda}=1.2\mum$ (panel (b)). The shaded regions mark the parameter space ($(a,U)$) where the rotational disruption occurs. One can see that $a_{\rm disr}$ decreases rapidly and $a_{\rm disr,max}$ increases linearly with the radiation strength $U$ (panels (a) and (b)). This reveals the range of the disruption size is broaden with increasing $U$. 

The disruption size decreases rapidly with increasing $U$ and then slowly approaches $a_{\rm c}$ when the centrifugal stress on the interface between the core and the mantle decreases to zero. The latter determines the lower boundary of rotational desorption determined by the strong grain core.

\begin{figure*}
\includegraphics[scale=0.5]{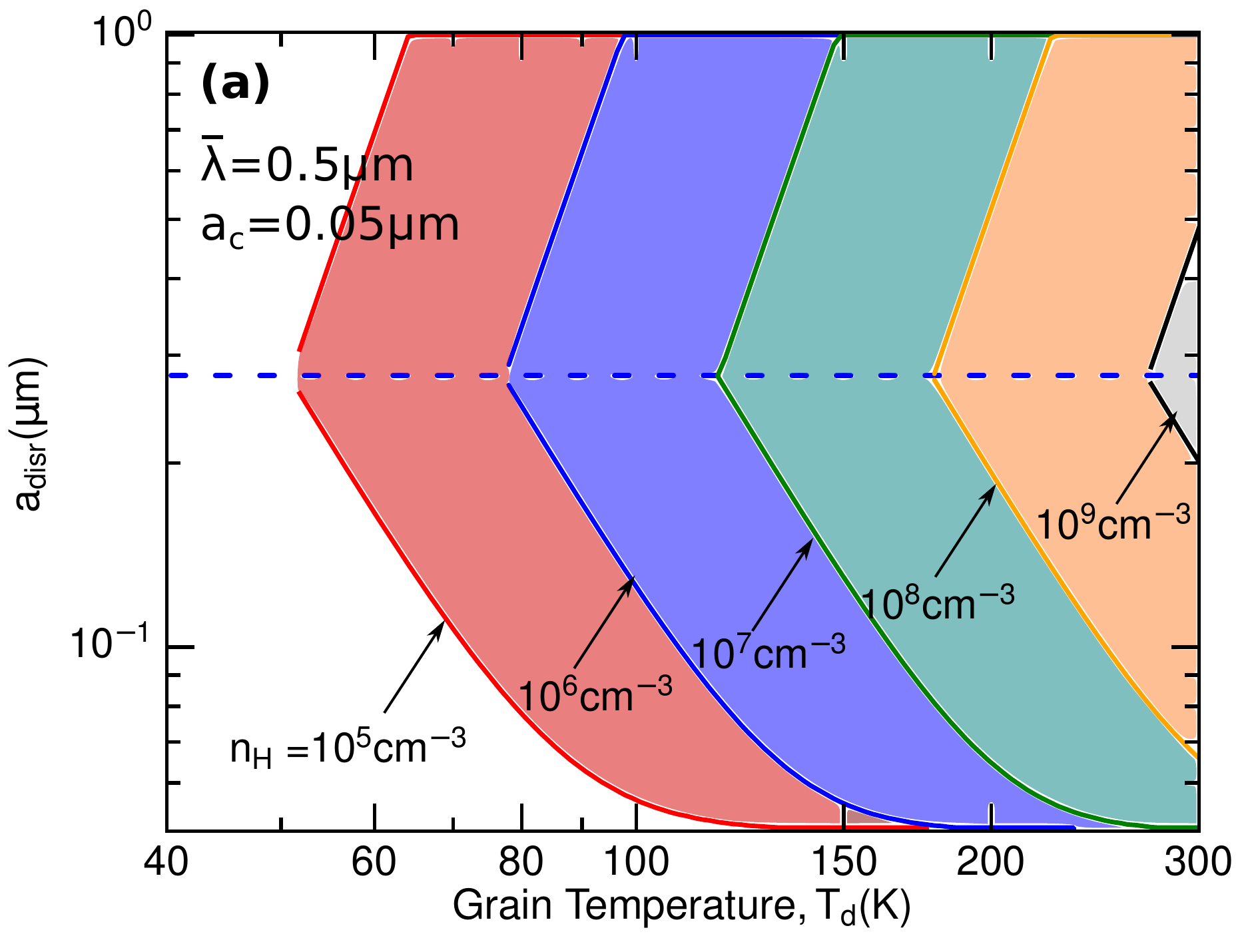}
\includegraphics[scale=0.5]{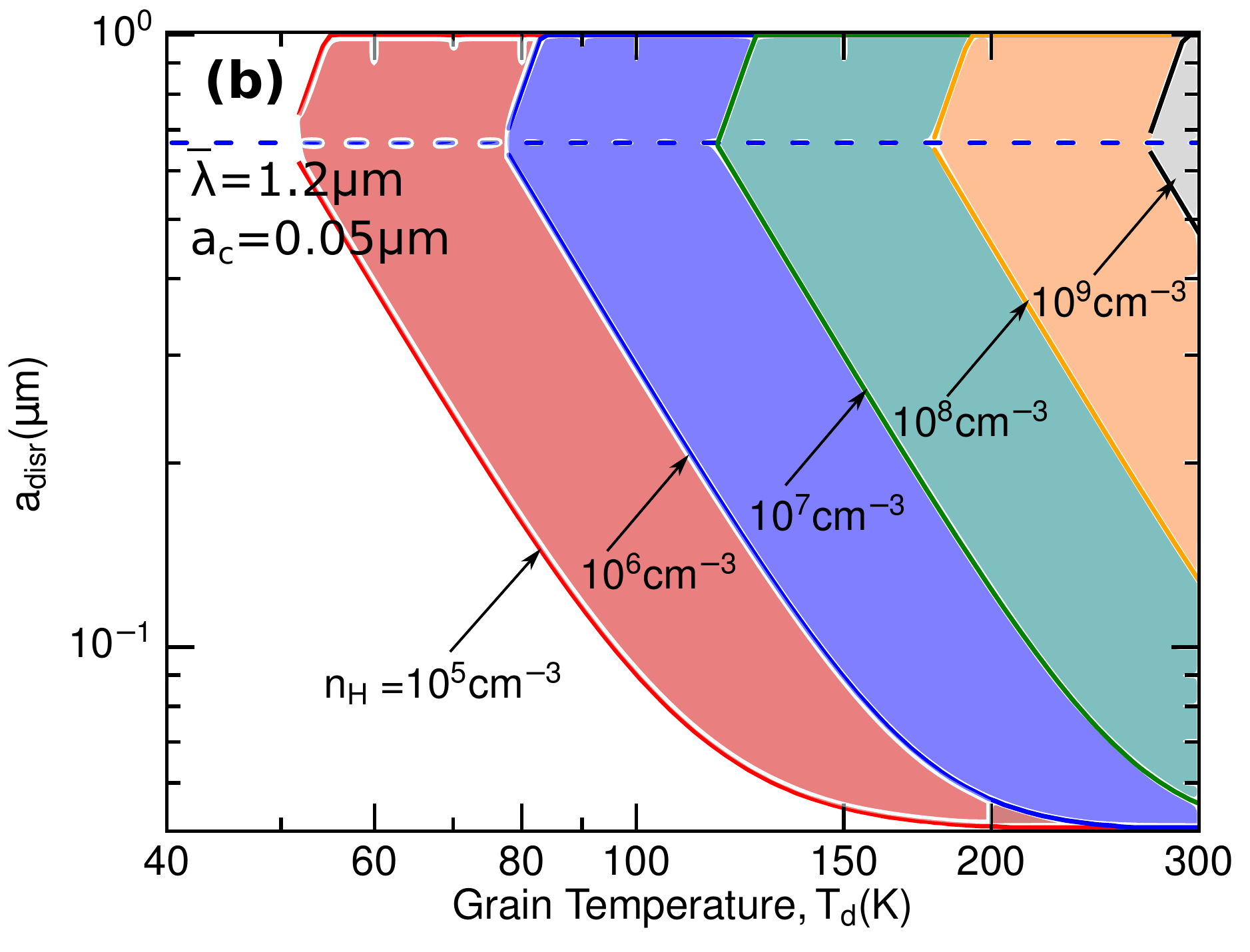}
\caption{Same as Figure \ref{fig:adisr_U_coreshell}, but for the dependence of grain temperature. At density of $n_{\H}\sim 10^{5}\cm^{-1}$, the ice mantle is removed at $T_{d}\gtrsim 60\K$, but at higher density $n_{\H}\sim 10^{6}\cm^{-3}$, the ice mantle is substantially removed only when $T_{d}\gtrsim 100\K$.}
\label{fig:adisr_Td_coreshell}
\end{figure*}

\begin{figure*}
\includegraphics[scale=0.5]{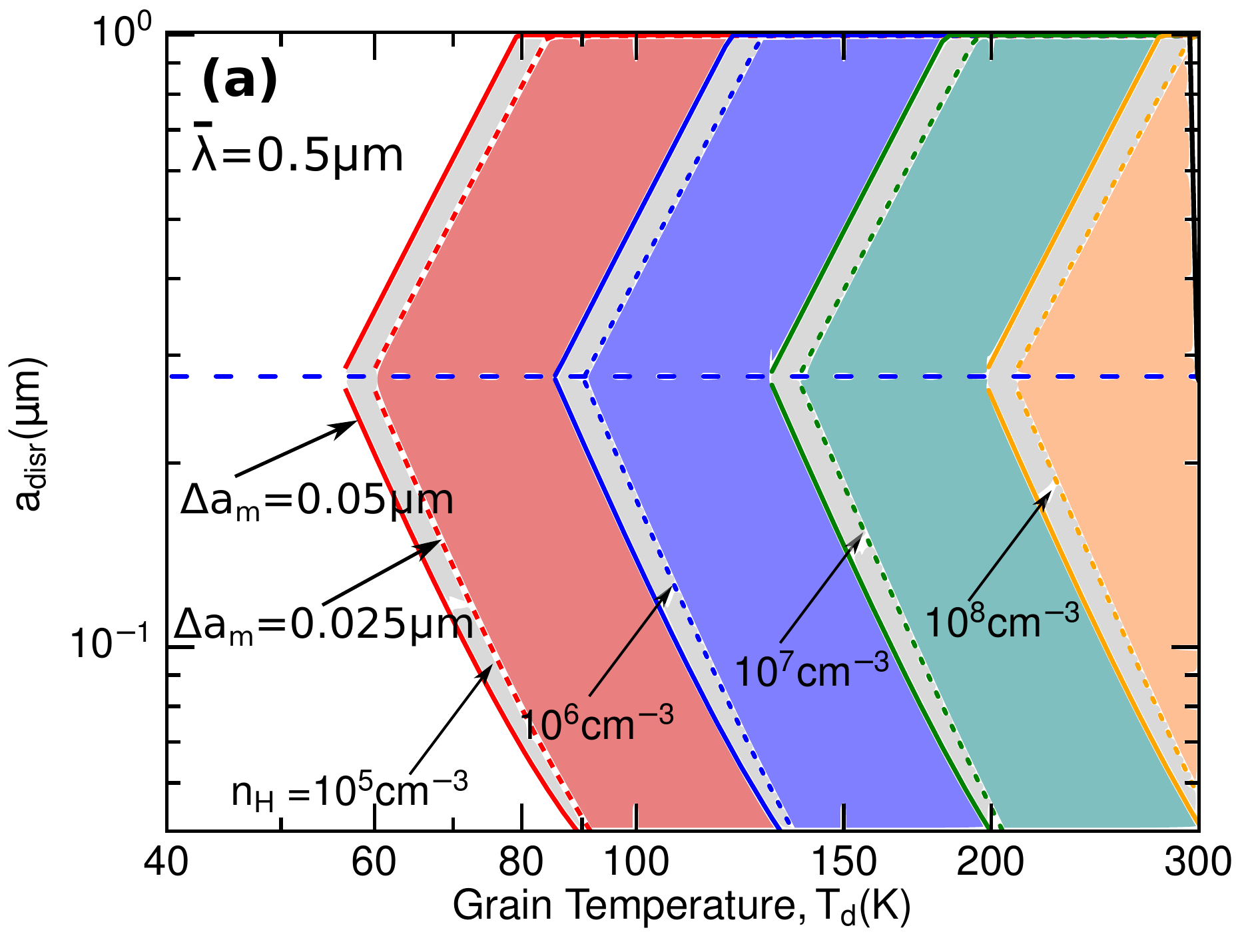}
\includegraphics[scale=0.5]{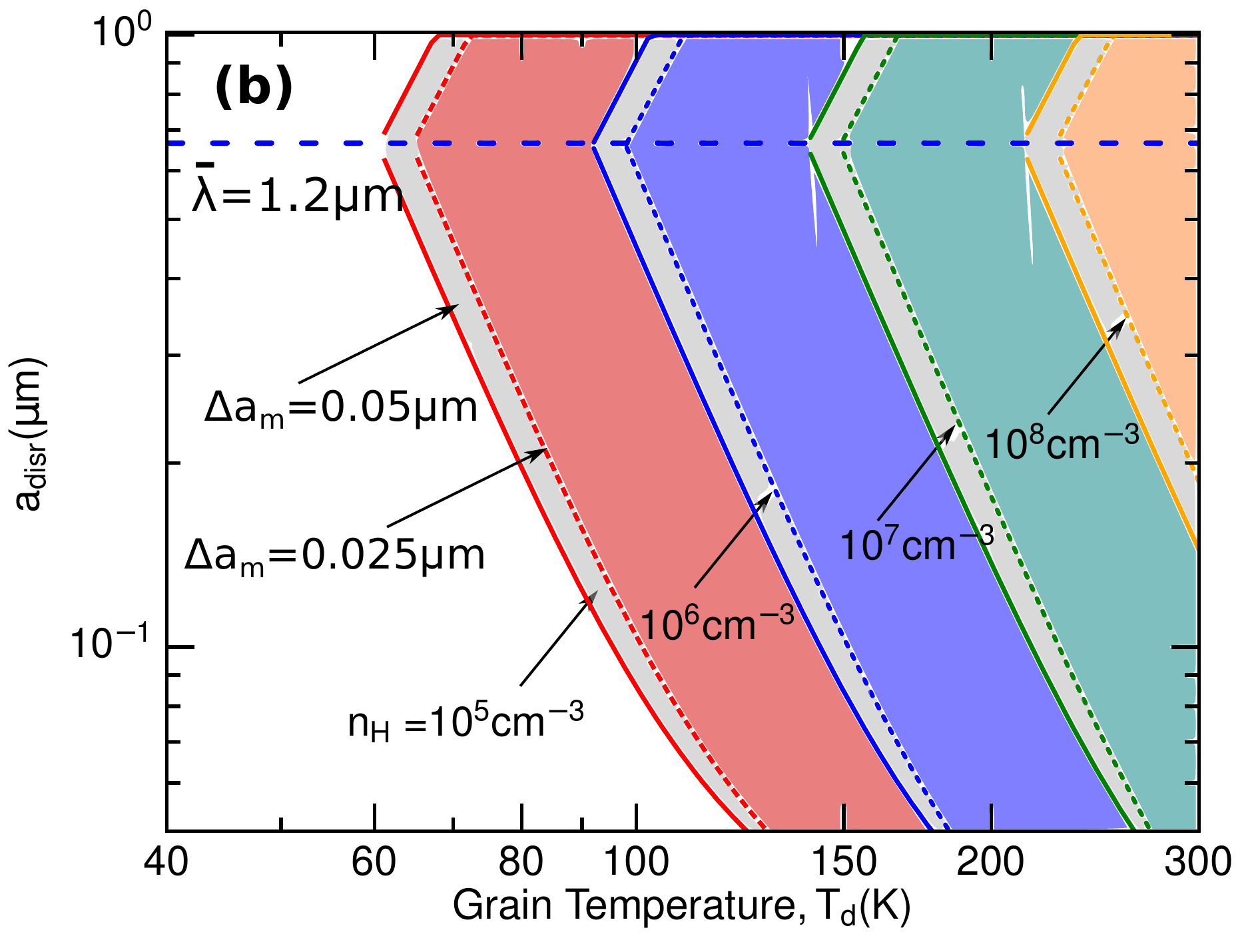}
\caption{Same as Figure \ref{fig:adisr_Td_coreshell}, but for the model with a fixed mantle thickness of $\Delta a_{m}=0.05\mum$ (solid lines) and $0.025\mum$ (dashed lines). For $n_{\H}\sim 10^{5}\cm^{-1}$, the ice mantle is removed at $T_{d}\gtrsim 60\K$, but for higher density $n_{\H}\sim 10^{6}\cm^{-3}$, the ice mantle is removed only when $T_{d}\gtrsim 100\K$.}
\label{fig:adisr_Td_coreshell_fixmantle}
\end{figure*}

We now assume that icy mantles of the same thickness cover all grain cores of the different sizes ($a_{c}$) and adopt the thickness $\Delta a_{m}=0.05\mum$. Results are shown in Figure \ref{fig:adisr_Td_coreshell_fixmantle}. We find that the disruption size is essentially the same for $a>a_{\rm c}$. The results are similar to those in Figure \ref{fig:adisr_U_coreshell} and (\ref{fig:adisr_Td_coreshell}), but the lower boundary is shifted to the corresponding value of $a_{c}$. Furthermore, the ice mantles on larger grain cores are desorbed at lower temperatures than from smaller grain cores, which originates from the increase of RATs with the grain size.

To easily compare rotational desorption with sublimation, in  Figure \ref{fig:adisr_Td_coreshell}, we show the desorption sizes as a function of the grain temperature where
\bea
T_{d}\simeq 16.4 a_{-5}^{-1/15}U^{1/6}\K \label{eq:Td}
\ena
for silicate grains (\citealt{2011piim.book.....D}).

For dense regions of density of $n_{\H}\sim 10^{5}-10^{6}\cm^{-3}$, a thick mantle of $a\sim 1\mum$ can be removed first at $T_{d}=80\K$. When the temperature increases to 100 K, the mantle layer has been removed until the grain radius is reduced to $\sim 0.1\mum$ can be removed. For higher density of $n_{\H}\sim 10^{8}\cm^{-3}$, the ice mantle starts to be separated at $T_{d}\sim 100\K$, and the entire mantle is completely removed at $T_{d}\sim 200\K$. For very dense regions of $n_{\H}\sim 10^{7}\cm^{-3}$, one see the ice mantle destruction is very efficient starting from $T_{d}\sim 100\K$. In extremely dense regions of $n_{\H}\sim 10^{9}\cm^{-3}$, stronger radiation intensity with $T_{d}\gtrsim 250\K$ can still disrupt ice mantles.

We now consider a core-ice mantle model in which the mantle thickness is fixed to $\Delta a_{m}=0.025\mum$ and $0.05\mum$, and the grain core radius $r_{c}$ is varied. Results are shown in Figure \ref{fig:adisr_Td_coreshell_fixmantle}. The desorption of the thicker mantle tends to occur at lower grain temperatures. Compared to Figure \ref{fig:adisr_Td_coreshell}, one can see that the ice mantle desorption requires higher radiation strength/temperature to be efficient. This feature originates from the fact that thicker ice mantles induce larger tensile stress acting on the interface between the mantle and the grain core (see Eq. \ref{eq:Sx}), which decreases the critical rotation rate and then requires a lower strength of ambient radiation fields. 

\subsection{Rotational desorption time vs. sublimation time}
\begin{figure*}
\includegraphics[scale=0.5]{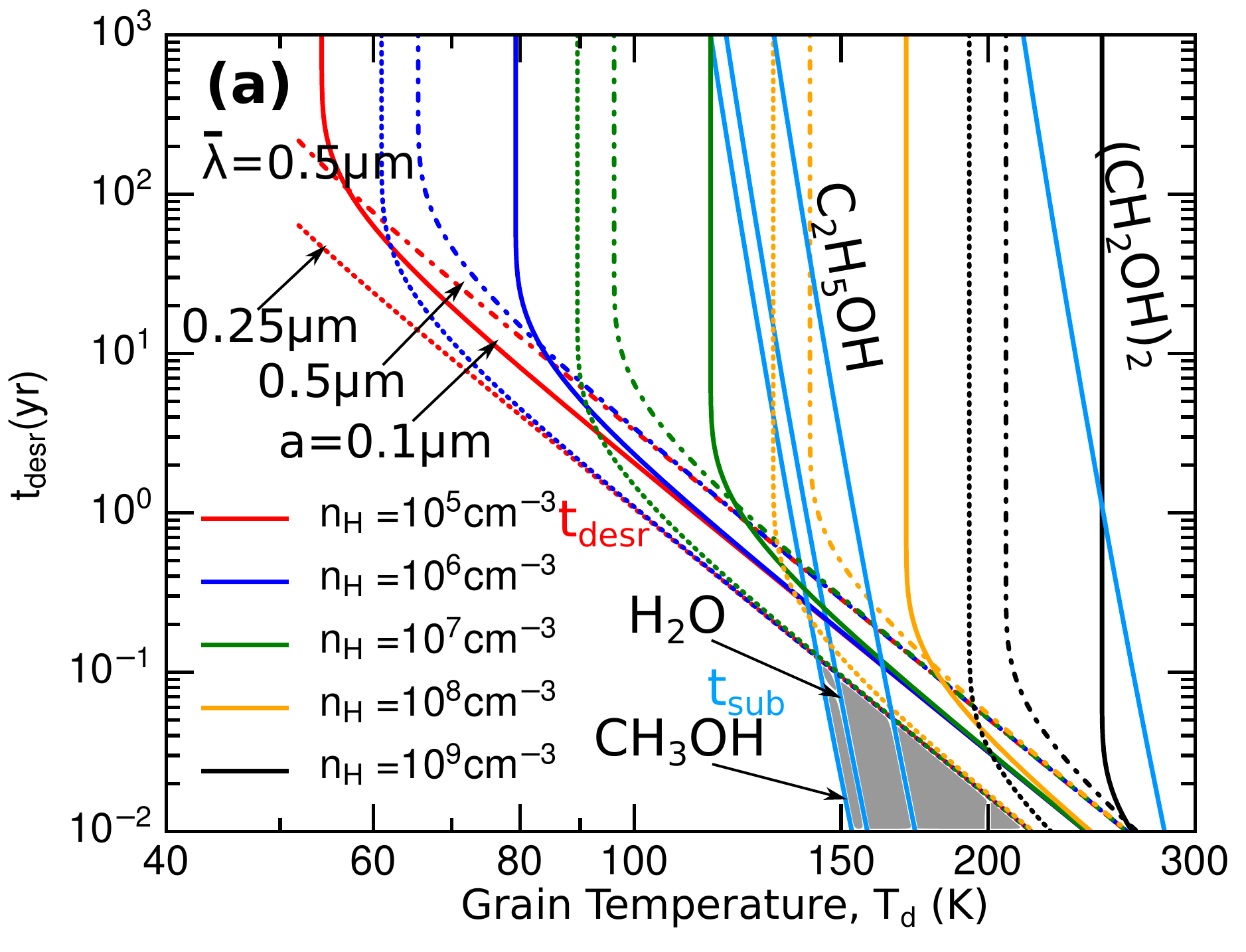}
\includegraphics[scale=0.5]{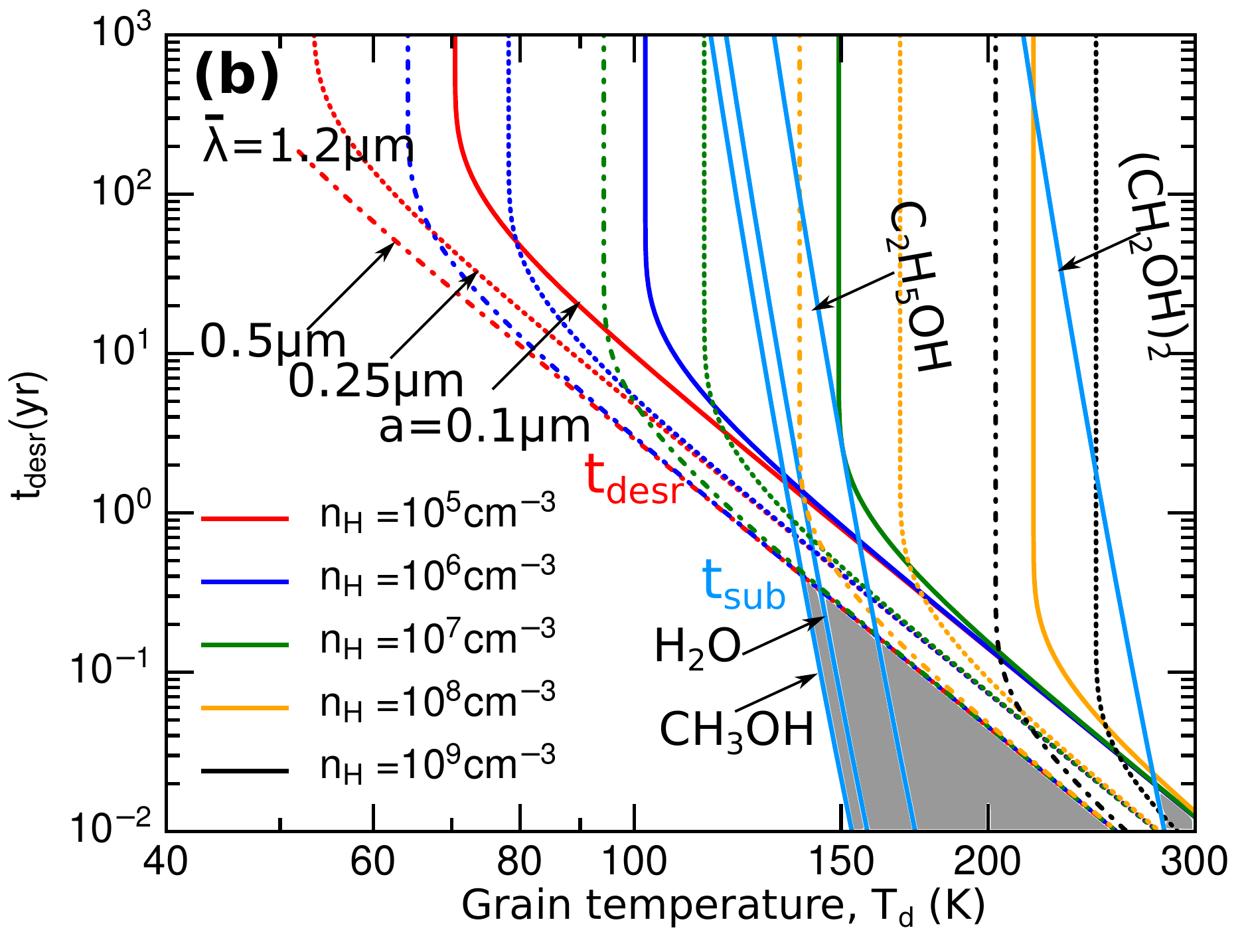}
\caption{Desorption time vs. grain temperature for the different grain sizes ($a=0.1, 0.25,0.5 \mum$). Thermal sublimation time of several popular COMs and water for a mantle thickness $\Delta a_{m}=0.025\mum$ is shown for comparison. Rotational desorption occurs faster than thermal sublimation and requires lower grain temperatures. Gray shaded areas mark the region where thermal sublimation is faster.}
\label{fig:tdisr_Td_coreshell}
\end{figure*}

We calculate the desorption time of ice mantles for the different grain temperatures, assuming the different grain sizes and a fixed mantle thickness of $\Delta a_{m}=0.05\mum$. Obtained results are shown in Figure \ref{fig:tdisr_Td_coreshell}.

For comparison, we also compute the sublimation time of the ice mantle as given by
\bea
t_{\rm sub}(T_d)=-\frac{\Delta a_{m}}{da/dt}
= \frac{\Delta a_{m}}{l\nu_{0}}\exp\left(\frac{E_{b}}{T_d}\right),\label{eq:tausub}
\ena
where $da/dt=l/\tau_{\rm evap}$ is the rate of decrease in the mantle thickness due to thermal sublimation, $l$ is the thickness of the ice monolayer, and $\tau_{\rm evap}$ is the characteristic time that molecules stay on the grain surface before evaporation:
\bea
\tau_{\rm evap}^{-1}=\nu_{0}\exp\left(\frac{-E_{b}}{T_d}\right),\label{eq:tevap}
\ena
where $\nu_{0}$ is the characteristic vibration frequency of the lattice, and $E_{b}$ is the binding energy (\citealt{1972ApJ...174..321W}). Table \ref{tab:Ebind} lists the binding energy and sublimation temperatures from experiments of the different molecules.

Plugging the numerical parameters of water ice into the above equation, we obtain
\bea
t_{\rm sub}\sim 1.5\times 10^{3}\left(\frac{\Delta a_{m}}{500\AAt}\right)\exp\left(\frac{E_{b}}{4800\K}\frac{100\K}{T_{d}} \right) \yr.\label{eq:tsub}
\ena

Comparing Equations (\ref{eq:tsub}) with (\ref{eq:tdisr}) one can see that rotational desorption is much faster than thermal sublimation of water ice at $T_{d}\sim 100\K$.

\begin{table}
\begin{center}
\caption{Binding energies and sublimation temperatures for selected molecules on an ice surface}\label{tab:Ebind}
\begin{tabular}{l l l} \hline\hline
{Molecules} & {$E_{b}$ (K)$^a$} & {$T_{\rm sub}$ (K)}\cr
\hline\\

$\rm H_{2}O$ & 5700 & 152$^b$ \cr
$\rm CH_{3}OH$ & 5530 & 99$^b$ \cr
$\rm HCOOH$ & 5570 & 155$^c$ \cr
$\rm CH_{3}CHO$ & 2775 & 30$^c$ \cr
$\rm C_{2}H_{5}OH$ & 6260 & 250$^c$ \cr
$\rm (CH_{2}OH)_2$ & 10200 & 350$^c$ \cr
$\rm NH_{3}$ & 5530 & 78$^b$\cr
$\rm CO_{2}$ & 2575 & 72$^b$ \cr
$\rm H_{2}CO$ & 2050 & 64$^b$ \cr
$\rm CH_{4}$ & 1300 & 31$^b$ \cr
$\rm CO$ & 1150 & 25$^b$ \cr

\cr
\hline
\multicolumn{3}{l}{$^a$~See Table 4 in \cite{2013ApJ...765...60G}}\cr
\multicolumn{3}{l}{$^b$~See Table 1 from \cite{1993prpl.conf.1177M}}\cr
\multicolumn{3}{l}{$^c$~See \cite{2004MNRAS.354.1133C}}\cr
\cr
\end{tabular}
\end{center}
\end{table}

Results for $t_{\rm sub}$ of several molecules with large binding energies, including H$_2$O, CH$_3$OH, HCOOH, C$_2$H$_5$OH, and (CH$_{2}$OH)$_{2}$, are shown in Figure \ref{fig:tdisr_Td_coreshell}. We adopt the characteristic frequency $\nu_{0}=5\times 10^{12}\s^{-1}$ (see e.g., \citealt{1993MNRAS.261...83H}). The disruption time is smaller for larger grain sizes, but it increases when $a> a_{\rm trans}$. Below the sublimation temperature (i.e., temperatures of $50-100\K$), rotational desorption is faster than sublimation for densities of $n_{\H}\lesssim 10^{6}\cm^{-3}$. At densities of $n_{\H}\sim 10^{7}\cm^{-3}$, rotational desorption is faster than sublimation of water ice at $T\sim 100-120\K$ (i.e., below sublimation limit of water ice). At higher density $n_{\H}\sim 10^{8}\cm^{-3}$, rotational desorption is faster than sublimation of ethanol for $T\sim 130-160\K$.

\subsection{Rapid evaporation of COMs and water ice from tiny fragments}
Below we will study the subsequent evaporation of COMs and water ice from tiny fragments produced by rotational desorption of ice mantles.

A detailed study about the size distribution of fragments resulting from rotational desorption of ice mantles is beyond the scope of this paper. Nevertheless, we can assume that resulting fragments include nanoparticles and tiny clusters of molecules (i.e., very small icy grains-VSG). The maximum size of fragments is perhaps comparable to the thickness of ice mantle of $\sim 25$ nm (i.e., 250 \AAt). The remaining question is how rapidly COMs and water ice can evaporate from such icy fragments?

Since icy fragments are exposed to the same strong stellar radiation as original large grains, due to low heat capacity, they can be rapidly heated to higher temperatures than the original grain (see \citealt{2001ApJ...551..807D} for more details). Indeed, the absorption of a single UV photon can instantaneously raise the temperature of the fragment to:
\bea
T_{\rm VSG} = \frac{\Delta E}{C_\V},\label{eq:TVSG}
\ena
where $\Delta E$ is the energy that an UV photon transfers to the dust grain, $C_\V=3N_{at}k_B$ is the volume heat capacity of the ice fragment, and $N_{at}=4/3 \pi a^{3} n_{\rm ice}$ is the total number of atoms. The number density of atoms the pure water ice mantle is $n_{\rm ice}\sim \rho_{\rm ice}/m(\rm H_{2}O)\approx 3.3\times 10^{22}\cm^{-3}$. 

The temperature of icy fragments can be rewritten as
\bea \label{eq:Td_nano}
T_{\rm VSG} \simeq 276 \left(\frac{E}{10\rm eV}\right) \left(\frac{a}{1\rm nm} \right)^{-3} \K,\label{eq:TVSG}
\ena 

It follows that for $a\lesssim 1$ nm, single-photon absorption can transiently heat the fragment to $T_{\rm VSG}\gtrsim 270\K$, which exceeds thermal sublimation threshold of COMs (see Table \ref{tab:Ebind}). In general, one expects a fraction of tiny fragments below $1$ nm, therefore, COMs and water ice can rapidly evaporate following the rotational desorption of ice mantles. 

Larger fragments of $a\sim 1-25$ nm can achieve equilibrium temperatures due to radiative heating, and their temperatures are considerably higher than that of original large grains of $a\sim 0.1\mum$, as given by Equation (\ref{eq:Td}). Due to a steep dependence of the sublimation rate on the grain temperature, a moderate increase in the temperature of small fragments can significantly increase the sublimation rate of COMs from these small fragments. Indeed, for $U\sim 10^{5}$, one has $T_{d}\sim 111.7\K$ for $a=0.1\mum$, and $T_{d}\sim 130.2\K$ for $a=0.01\mum$. The ratio of their sublimation rates is equal to 
\bea
\frac{\tau_{\rm sub}(a=0.01\mum)^{-1}}{\tau_{\rm sub}(a=0.1\mum)^{-1}}\sim \frac{\exp(-4800/130.2)}{\exp(-4800/111.7)}\sim 450.\label{eq:tsub_frag}
\ena

\begin{figure}
\includegraphics[scale=0.45]{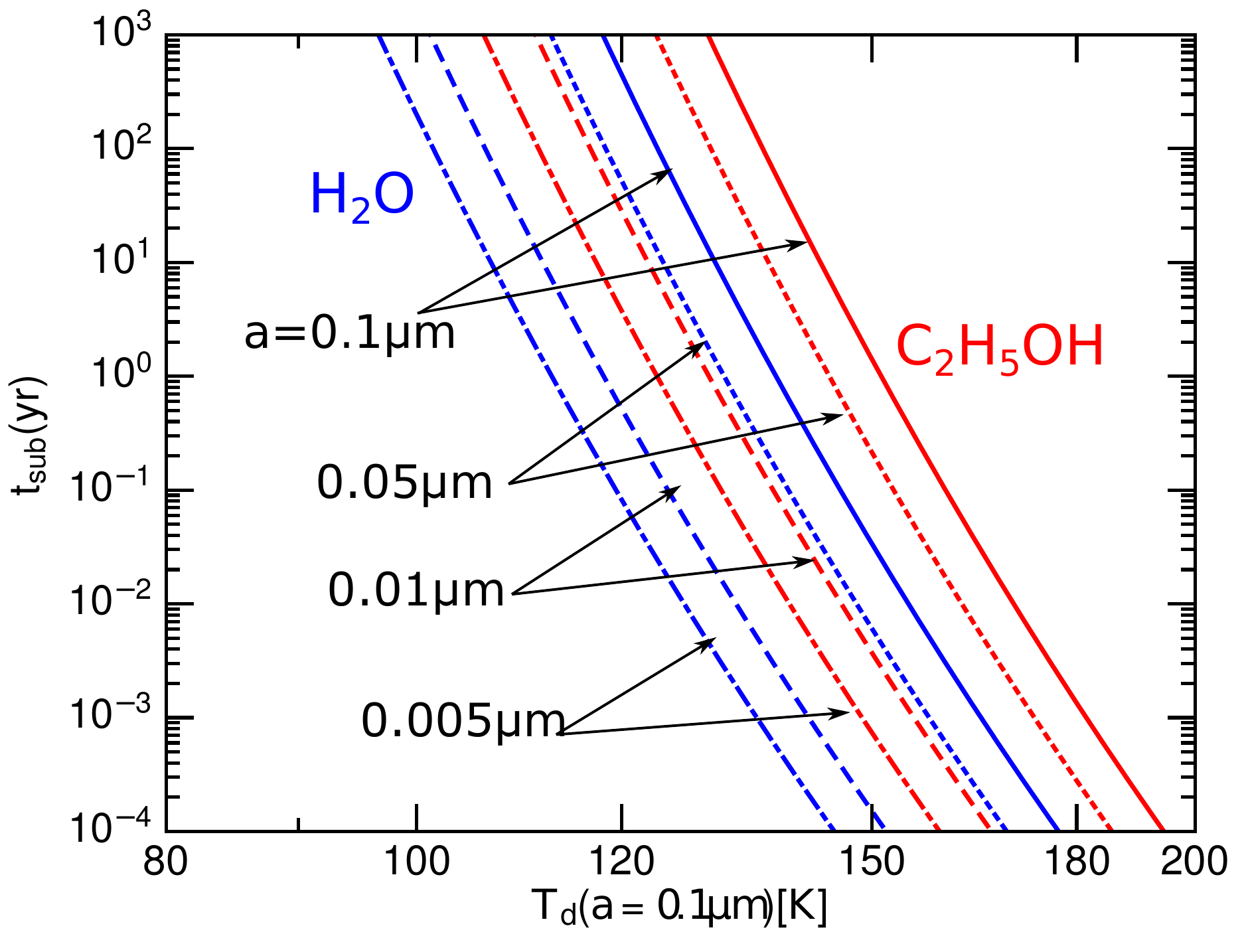}
\includegraphics[scale=0.45]{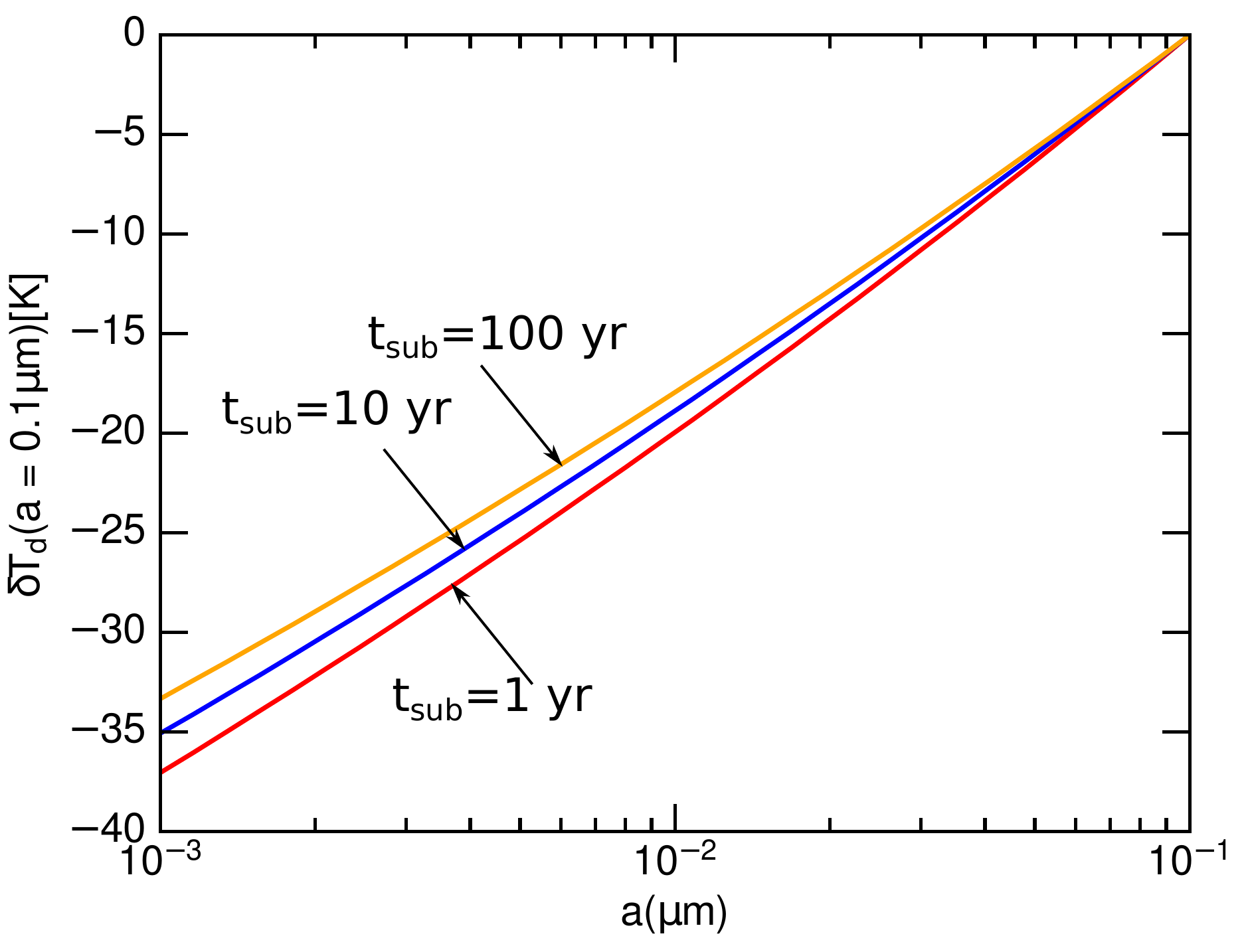}
\caption{Upper panel: thermal sublimation time of water ice and ethanol vs. the temperature of $0.1\mum$ grains for the different sizes. The sublimation time is significantly decreased with decreasing the grain radius. Lower panel: the decrease of grain temperature required to produce the same sublimation rate vs. grain size, assuming $t_{\rm sub}=1, 10, 100$ yr.}
\label{fig:tsub_Td}
\end{figure}

Figure \ref{fig:tsub_Td} (upper panel) shows the sublimation time for three different grain sizes as a function of the temperature of $a=0.1\mum$ grains. For a given temperature, the sublimation rate of smaller grains is several orders of magnitude larger than that of large grains. Therefore, by disrupting the ice mantle on a large grain into small fragments, rotational desorption can allow the sublimation of COMs and water ice at much faster rate compared to from the original icy grain mantle. 

Figure \ref{fig:tsub_Td} (lower panel) shows the decrease of grain temperature to have the same sublimation rate as a function of grain sizes. The decrease of grain temperature decreases rapidly with the fragment size and achieves $|\delta T|\gtrsim 20\K$ for $a\lesssim0.01\mum$. 

\begin{figure*}
\includegraphics[scale=0.5]{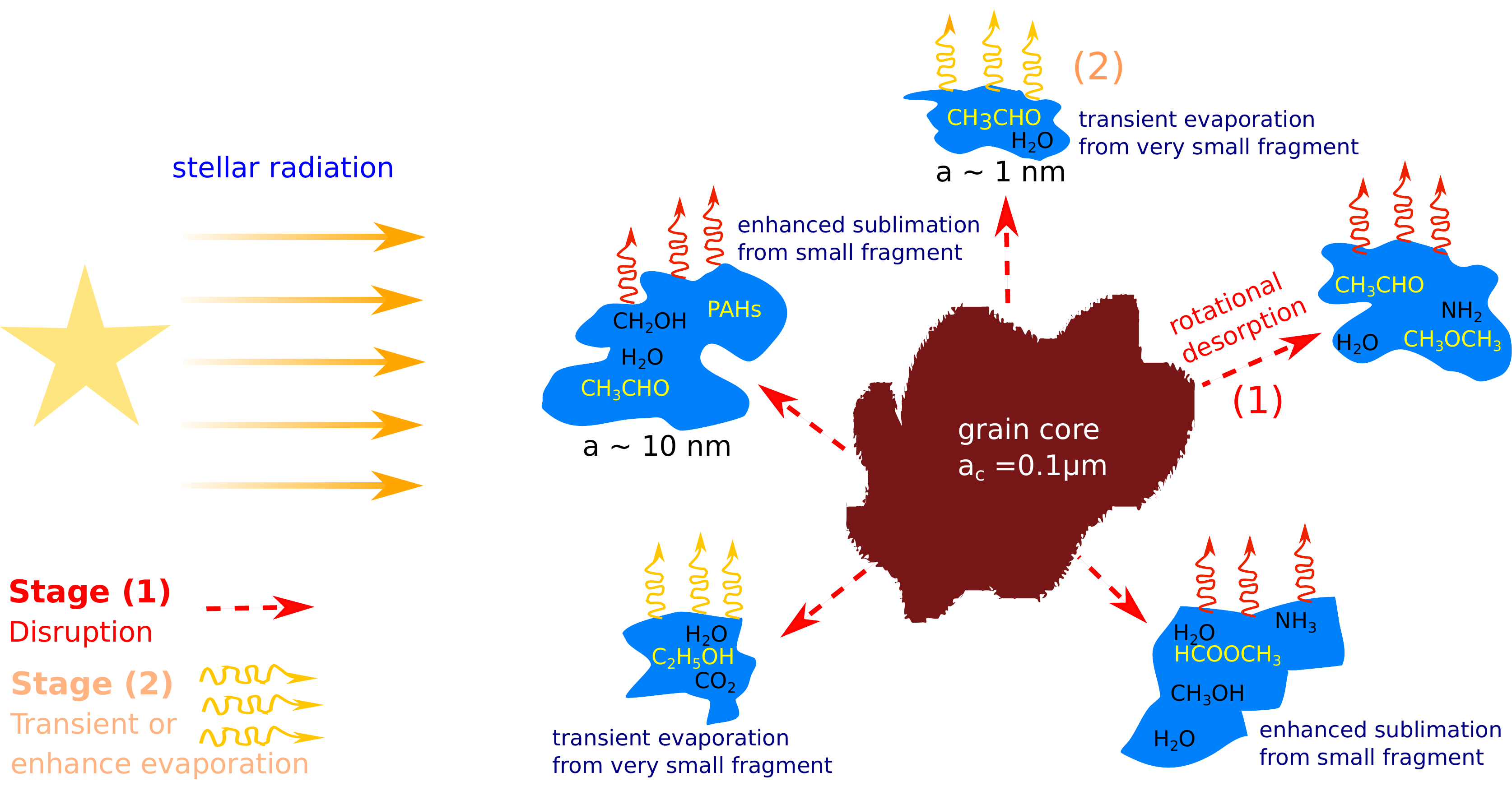}
\caption{Schematic illustration of rotational desorption process of COMs from icy grain mantles comprising two stages: (1) disruption of icy mantes into small fragments by RATD, and (2) rapid evaporation of COMs due to thermal spikes for very small fragments or increased sublimation for larger fragments.}
\label{fig:rotdesr}
\end{figure*}

To summarize the rotational desorption mechanism introduced in this section, in Figure \ref{fig:rotdesr}, we illustrate the rotational desorption process of COMs from icy grain mantles which includes two stages. Firstly, the ice mantle of a large core-ice mantle grain ($a_{c}=0.1\mum$) is disrupted into small fragments by means of RATD. Secondly, very small fragments ($a\lesssim 1$ nm) are transiently heated to high temperatures, inducing transient evaporation of molecules. Larger fragments (e.g., $a\sim 1-10$ nm) can be heated to higher temperatures than the original grain, which significantly enhances the rate of thermal sublimation as shown in Figure \ref{fig:tsub_Td}.

\section{Application: Rotational desorption of ice mantles in hot cores/corinos}\label{sec:hotcore}

We now apply our theory in the previous section to study the desorption of ice mantles from grains in hot cores and hot corinos, which are inner regions surrounding high-mass protostars (\citealt{1987ApJ...315..621B}) and low-mass protostars (\citealt{2004ApJ...615..354B}; \citealt{2018arXiv180608137B}).

Hot cores/corinos are directly heated by the central protostar. Let $L$ be the bolometric luminosity of the protostar. The radiation strength at distance $r$ from the source is given by
\bea
U(r)=\left(\frac{L}{4\pi r^{2}c u_{\rm ISRF}}\right)=U_{\rm in}\left(\frac{r_{\rm in}}{r}\right)^{2},
\ena
where $U_{\rm in}$ denotes the radiation strength at inner radius $r_{\rm in}$. 

The gas density and temperature can be approximately described by power laws:
\bea
n_{\gas}=n_{\rm in}\left(\frac{r_{\rm in}}{r}\right)^{p},\\
T_{\gas}=T_{\rm in}\left(\frac{r_{\rm in}}{r}\right)^{q},
\ena
where $n_{\rm in}$ and $T_{\rm in}$ are gas density and temperature at radius $r_{\rm in}$, and $q=2/(4+\beta)$ with $\beta$ the dust opacity index (see e.g., \citealt{2000ApJ...530..851C}). The typical density profile in the inner hot region is $p\sim 1.5$. A more detailed model of hot cores is presented in \cite{Nomura:2004jt}.

From Equation (\ref{eq:adisr_low}), one obtains the disruption size of ice mantles as follows
\bea
a_{\rm disr}(r)&\simeq&0.13\gamma^{-1/1.7}\bar{\lambda}_
{0.5}(S_{\max,7}/\hat{\rho}_{\rm ice})^{1/3.4}(1+F_{\rm IR})^{1/1.7}\nonumber\\
&\times&\left(\frac{n_{\rm in}T_{\rm in}^{1/2}}{100U_{\rm in}}\right)^{1/1.7}\left(\frac{r_{\rm in}}{r}\right)^{(p+q/2-2)/1.7}\mum,~~~\label{eq:adisr_core}
\ena
which slowly decreases with radius $r$ as $r^{(p+q/2-2)/1.7}\sim r^{-0.1}$ for typical slopes.

For low-mass protostars, one can assume $r_{\rm in}=25\AU$ and $n_{\rm in}\sim 10^{8}\cm^{-3}$ and $L=36L_{\odot}$ \citep{Visser:2012km}, one gets $a_{\rm disr}\sim 0.29\mum$ at $r=r_{\rm in}$ and $a_{\rm disr}\sim 0.63\mum$ for $r=10r_{\rm in}$. For hot cores, we adopt a typical luminosity of $L=10^{5}L_{\odot}$ and typical parameters $r_{\rm in}\sim 500 \AU, n_{\rm in}\sim 10^{8}\cm^{-3}, U_{\rm in}\sim 2\times 10^{7}$ and $T_{\rm in}\sim 274\K$ (see e.g., \citealt{Bisschop:2007cu}). Therefore, Equation (\ref{eq:adisr_core}) gives $a_{\rm disr}=0.1\mum$ and $0.16\mum$ at $r=r_{\rm in}, 10r_{\rm in}$ respectively. The results for $n_{\rm in}=10^{7}\cm^{-3}$ as usually assumed \citep{1999MNRAS.305..755V} are even more promising. 

\begin{figure}
\includegraphics[scale=0.45]{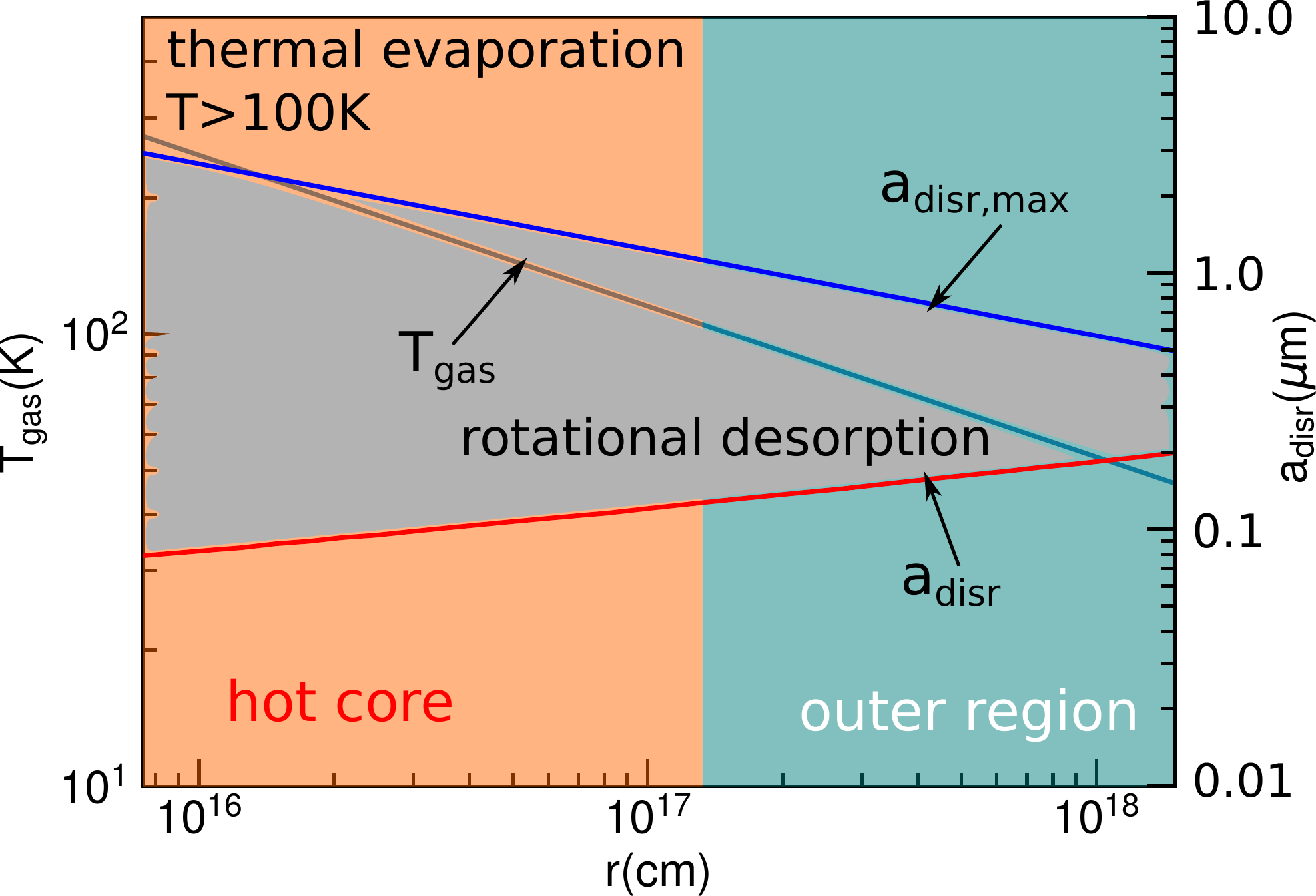}
\caption{Gas temperature and grain disruption size vs. radius for an envelope around a high-mass protostar. Classical thermal sublimation is important only for the inner region (hot core), whereas rotational desorption is important for both hot core and outer region.}
\label{fig:adisr_Tn}
\end{figure}

Figure \ref{fig:adisr_Tn} illustrates the importance of rotational desorption vs. classical thermal sublimation of ice mantles around a protostar of $L=10^{5}L_{\odot}$ and $\bar{\lambda}=0.5\mum$. Thermal evaporation is important only in the inner regions where $T_{\gas}>100\K$, whereas rotational desorption can be efficient at larger radii with low temperatures of $T_{\gas}\sim 40-100\K$. 

We note that even in the hot inner region where thermal sublimation is active, rotational desorption and ro-thermal desorption (\citealt{Hoang:2019ij}) are more efficient than the classical sublimation for molecules with high binding energy such as water and COMs.

The efficiency of rotational desorption in cold extended regions shown in Figure \ref{fig:adisr_Tn} successfully explain the presence of COMs from outer extended regions around hot cores by \cite{Fayolle:2015cu}. Furthermore, this mechanism can explain the presence of HCOOH, CH$_3$CHO from cold regions with $T_{\gas}\sim 40-70\K$ (below the sublimation threshold of these molecules) and low column density by \cite{Bisschop:2007cu}. Future high-resolution observations by ALMA would be unique to test our prediction of an extended regions of COMs.

\section{Discussion}\label{sec:discuss}

\subsection{Comparison of rotational desorption with other desorption mechanisms in star-forming regions}
First of all, we should stress that in this paper we are primarily concerned with desorption of molecules in star-forming regions where icy grains are illuminated by protostars and quantify an important physical effect, namely, suprathermal rotation of grains, that is disregarded in previous studies. 

Both rotational desorption and thermal sublimation of COMs from ice mantles rely on interaction of radiation fields with dust grains to be efficient. Here, we outline the key differences of the rotational desorption mechanism that can be tested with observations.

From Figure \ref{fig:tdisr_Td_coreshell} one can see that rotational desorption can occur at low temperatures below the sublimation threshold of COMs and water ice, and the critical temperature for rotational desorption decreases with decreasing the local gas density (\citealt{Hoang:2019da}; see also Section \ref{fig:tdisr_Td_coreshell}). It can be effective even at temperatures of $T<50\K$ if the local gas density is $n_{\H}<10^{5}\cm^{-3}$. Therefore, COMs/water vapor can be observed in cold regions ($T<100\K$) with not very high gas density. On the other hand, thermal sublimation only depends on the grain temperature which requires intense radiation such that grains can be heated to $T>100\K$. 

Second, compared to classical sublimation, the circumstellar region in which rotational desorption is important would be more extended due to lower required radiation intensity. For instance, using $U\sim T_{d}^{6}$, one can estimate the radius of regions in which rotational desorption is efficient as $R_{\rm desr}/R_{\rm sub}=(U_{\rm sub}/U_{\rm desr})^{1/2}\sim (T_{\rm sub}/T_{\rm desr})^{3}\sim (100/80)^{3}\sim 2$. Therefore, we predict that COMs can be released in a region more extended than classical sublimation by a factor of 2. 

Third, through rotational desorption, COMs and water ice can be released simultaneously from tiny fragments ($a<1$ nm). For larger fragments ($a\sim 1 nm-\Delta a_{m}$), the sublimation is essentially classical one but with significantly enhanced rates, and COMs of the different binding energies are released at the different distances from the radiation source.

Finally, while photodesorption requires far-UV photons ($h\nu\gtrsim 6$ eV or $\lambda<0.2\mum$) to be effective (\citealt{2009ApJ...693.1209O}), rotational desorption can work with a broad range of the radiation spectrum (i.e., $\lambda>0.2\mum$) because RATs depend on the ratio of the photon wavelength to the grain size (\citealt{2007MNRAS.378..910L}; \citealt{Herranen:2019kj}). Let us estimate the lifetime of an icy grain by UV photodesorption. Let $Y_{\rm pd}$ be the photodesorption  yield of water ice which is defined as the fraction of molecules ejected over the total number of incident UV photons. The rate of mass loss due to UV photodesorption is
\bea
\frac{dm}{dt} = \frac{4\pi a^{2} \rho_{ice} da}{dt}=\bar{m}Y_{\rm pd}F_{\rm FUV}\pi a^{2},
\ena
where $\bar{m}$ is the mean mass of ejected molecules, $F_{\rm FUV}$ is the flux of FUV photons. Let $G_{0}=F_{\rm FUV}/F_{\rm FUV,ISRF}$ be the strength of the UV photons relative to the standard interstellar radiation field. Then, the above equation becomes:
\bea
\frac{da}{dt} = 2.2\times 10^{-3}G_{0}\left(\frac{Y_{\rm pd}}{0.001}\right) \frac{\AAt}{\yr},
\ena
where $\bar{m}=m(\rm H_{2}O)$ and the typical yield $Y_{\rm pd}=0.001$ \citep{2009ApJ...693.1209O} is adopted.

Using the typical parameters for a hot core, one obtains the photodesorption time for an ice mantle of thickness $\Delta a_{m}$:
\bea
t_{\rm pd}&=&\frac{\Delta a_{m}}{da/dt}\simeq \left(\frac{\Delta a_{m}(\AAt)}{2.2\times 10^{-3}\AAt}\right)\left(\frac{10^{-3}}{Y_{\rm pd}}\right)\left(\frac{1}{G_{0}}\right)~ \yr\nonumber\\
&\simeq& 2.2\left(\frac{\Delta a_{m}}{500\AAt}\right) \left(\frac{10^{-3}}{Y_{\rm pd}}\right)\left(\frac{10^{5}}{G_{0}}\right)\yr.
\ena

Comparing $t_{\rm pd}$ with $t_{\rm disr}$ (Eq. \ref{eq:tdisr}) one can see that the rotational disruption is one order of magnitude faster than UV photodesorption, assuming $G_{0}\sim U_{0}\sim 10^{5}$ and $a=0.1\mum$. In the shielded region of optical depth $\tau_{V}$, the UV strength is reduced to $G=G_{0}e^{-\tau_{\rm FUV}}$, but the optical radiation is reduced to $U=U_{0}e^{-\tau_{V}}$ only where $\tau_{\rm FUV}\sim 2-3\tau_{V}$ (see e.g., \citealt{2001ApJ...548..296W}). The UV photodesorption yield of methanol is found to be very low of $<10^{-6}$ \citep{2016ApJ...817L..12B}, although this yield only take into account intact methanol molecules. One important aspect is that the penetration length of FUV photons is much shorter than optical photons due to higher dust extinction. As a result, rotational desorption is efficient in more extended regions  around YSOs than photodesorption.

\subsection{Rotational desorption enhances abundance of COMs in the outflow and the outflow cavity walls of protostars, and PDRs}
Due to the clearing-out by outflows, stellar radiation can freely propagate through the outflow cavity (of gas density $n_{\H}\sim 10^{3}\cm^{-3}$) and illuminate icy grains in the cavity wall (boundary region between outflow and envelope). The cavity wall behaves like a photo-dominated regions where the radiation strength can be large of $U\sim 10^{3}-10^{4}$, and the gas density of the outflow cavity wall is $n_{\H}\sim 10^{4}-10^{6}\cm^{-3}$ (\citealt{Visser:2012km}). With these physical parameters, using the results from Figure \ref{fig:adisr_Td_coreshell_fixmantle}, one can see that the entire mantle can be evaporated via rotational desorption, which increases the abundance of COMs in the cavity wall.

Rotational desorption of ice mantles is also expected to increase the abundance of COMs in the outflows of protostars because of high radiation intensity and low gas density. Indeed, observations usually show an enhancement of COMs in the outflows of low-mass protostars \cite{Drozdovskaya:2015jx} and high-mass protostar \citep{2017MNRAS.467.2723P}. COMs are also observed to be more abundant in the shock L1157-B1 than in hot corinos (\citealt{2017MNRAS.469L..73L}). We note that \cite{Drozdovskaya:2015jx} explained the enhancement of COMs by means of UV irradiation from protostars due to the outflow cavity (see also \citealt{Karska:2014bt}). Yet, an enhanced irradiation would increase the efficiency of photodestruction of COMs. \cite{2017MNRAS.467.2723P} explained this enhancement due to sputtering and grain-grain collisions in C-shocks of velocities $v<40\km \s^{-1}$ in the cavity walls. However, a recent work by \cite{Godard:2019kb} shows that C-shocks shrink in the radiation fields of $G_{0}> 0.2 (n_{\H}/1\cm^{-3})^{1/2}$, which would reduce the efficiency of sputtering. 

At far distances from the protostar, the effect of shocks can be important. As shown in \cite{2019ApJ...877...36H} and \cite{Le:2019wo}, C-shocks can spin-up nanoparticles to suprathermal rotation due to supersonic drift of neutrals relative to charged grains. Then icy grain mantles can be released, producing gas-phase COMs by either the rapid disruption of weak nanoparticles, or rotational desorption of strong nanoparticles \citep{Le:2019wo}.

Finally, rotation desorption of ice mantles is important for PDRs, which are dense ($n_{\H}\sim 10^{4}-10^{5}\cm^{-3}$) and directly illuminated by strong stellar radiation of $U\sim 10^{3}-10^{5}$. We note that outflow cavity walls and surface of circumstellar disk are also considered PDRs (\citealt{1995ApJ...455L.167S}). Thus, rotational desorption is expected to be efficient in releasing COMs from icy mantles for these PDR environments. Using the results in Figure \ref{fig:adisr_Td_coreshell_fixmantle}, one can see that the icy mantles can be rapidly removed from $a\gtrsim 0.2\mum$ grains. This mechanism can explain the detection of COMs and polycyclic aromatic hydrocarbons in PDRs (see \citealt{2017SSRv..212....1C} for a review).

\subsection{Effect of grain evolution on grain-surface chemistry: link between dust properties and COMs}
Our numerical results presented in Section \ref{sec:desorp_grain} are also obtained for core-ice mantle grains with the maximum size of $a_{\max}\sim 1\mum$, and the typical grain size adopted for chemical modeling of gas-grain chemistry is $a=0.1\mum$ (see e.g., \citealt{2013ApJ...765...60G}). However, in dense prestellar cores and PPDs, grain coagulation due to grain-grain collisions is expected to form large fluffy aggregates of original icy grains, namely composite grains (\citealt{1994A&A...291..943O}; \citealt{2013MNRAS.434L..70H}). Various observations show the signature of grain growth in prestellar cores (\citealt{2010Sci...329.1622P}) and young protostellar system \citep{2009ApJ...696..841K}. Note that, in dense molecular clouds where interstellar radiation is significantly attenuated, coagulation of icy grains can proceed  to form large aggregates, without being rotationally disrupted by RATD. However, when exposed to strong radiation of protostars/young stars, such large composite grains are expected to be destroyed via the RATD mechanism (\citealt{Hoang:2019da}). 

The disruption of large grains can also be calculated using the method in Section \ref{sec:desorp_grain}, but with the tensile strength of composite grains (\citealt{1995A&A...295L..35G}; \citealt{1997A&A...323..566L}; \citealt{2019ApJ...876...13H}). Experimental data show a wide range of the tensile strength for aggregate grains, from $10^{4}-10^{6}\erg\cm^{-3}$, depending on the particle radius \citep{Gundlach:2018cu}. Thus, for our calculations, we assume the tensile strength $S_{\max}\sim 10^{5}\erg\cm^{-3}$ which correspond the the average particle radius $a_{p}=10$ nm, assuming a porosity of $20\%$ (see \citealt{2019ApJ...876...13H}).  

\begin{figure*}
\includegraphics[scale=0.5]{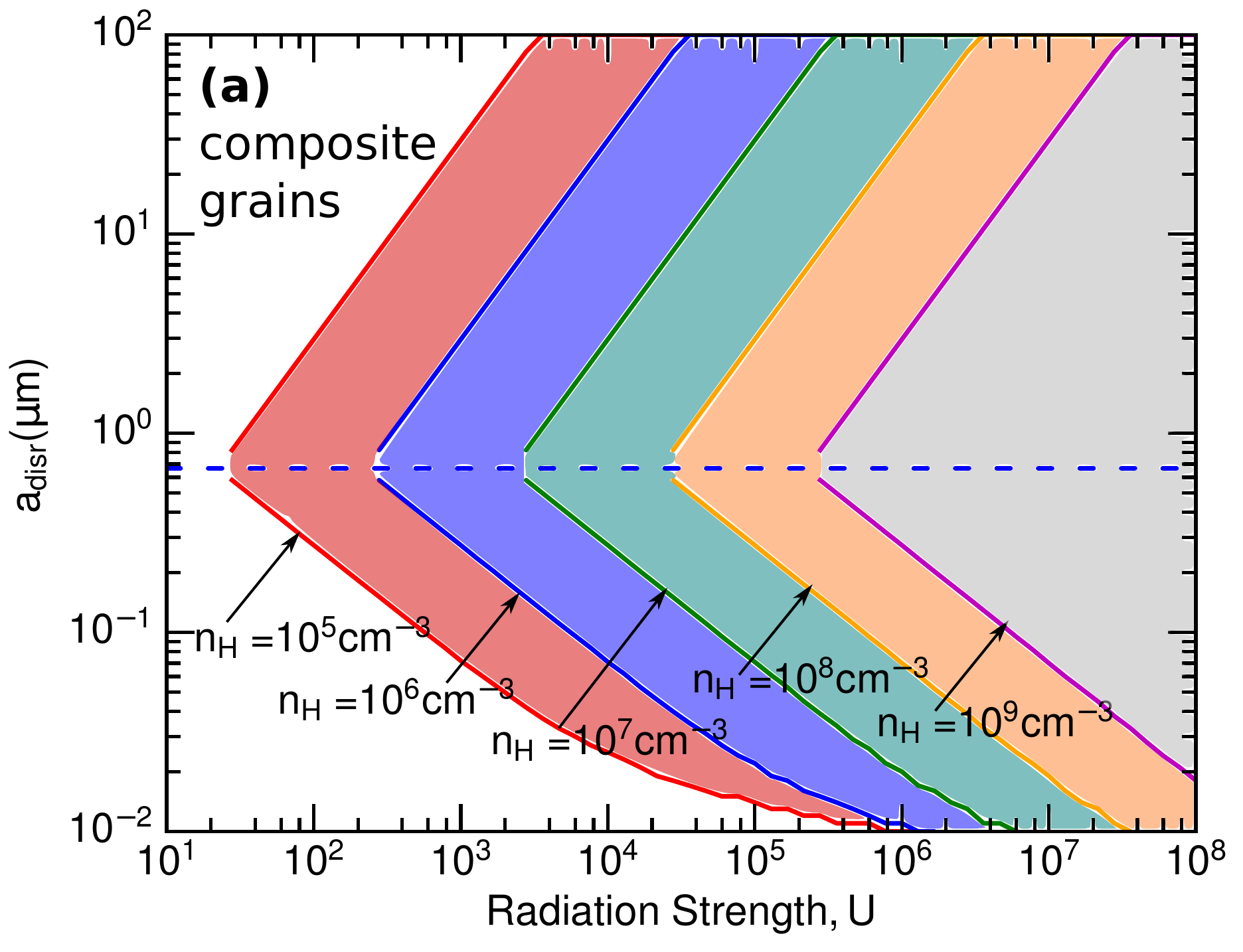}
\includegraphics[scale=0.5]{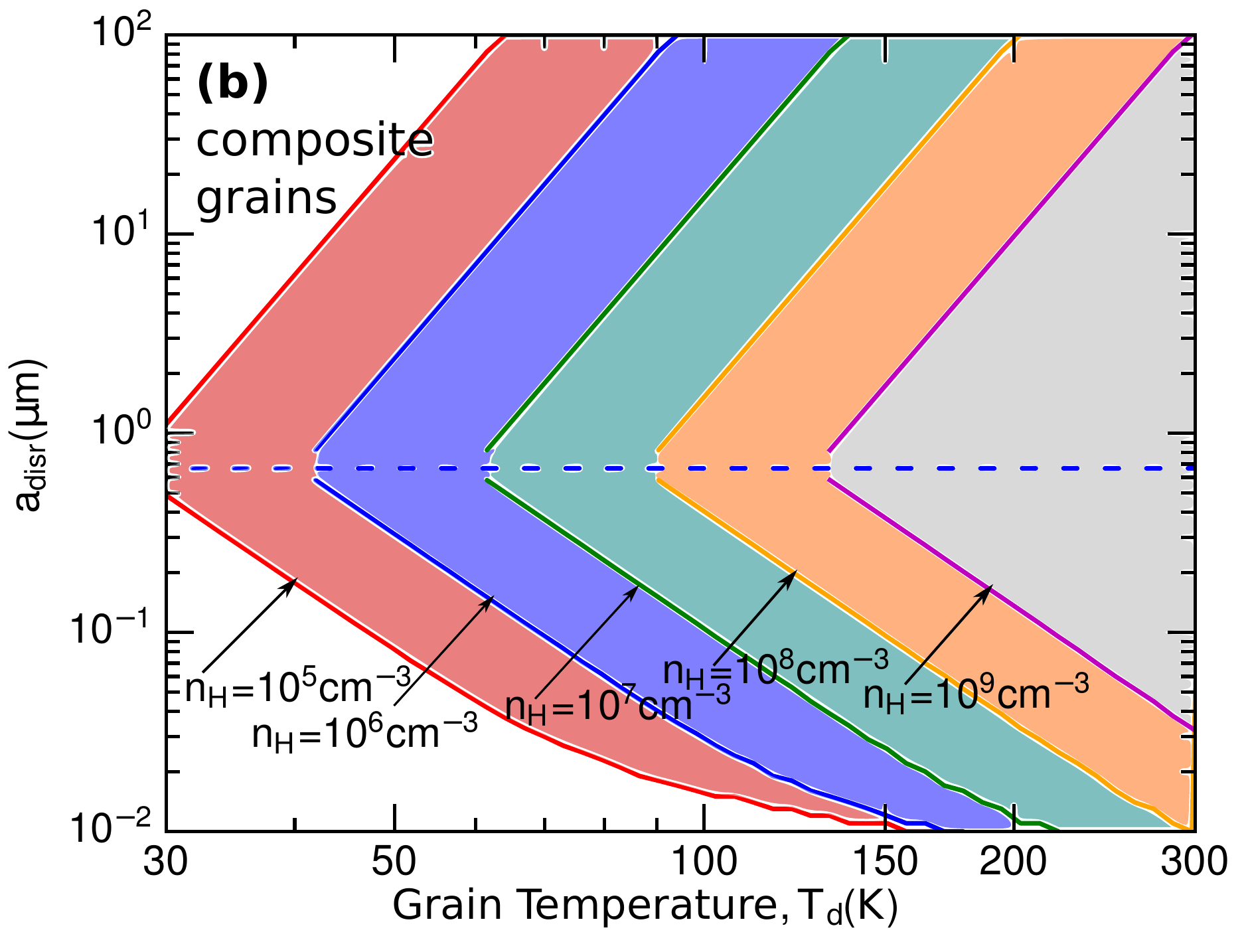}
\caption{Disruption sizes of composite grains as a function of radiation strength (left panel) and grain temperature estimated at $a=0.1\mum$ (right panel). Composite grains of $a\sim 0.3-3\mum$ can be disrupted at temperatures $T_{d}<50\K$ at density $n_{\H} \lesssim 10^{6}\cm^{-3}$, but for very dense and cold regions ($n_{\H}>10^{9}\cm^{-3}, T_{d}<100\K$) very large grains can survive.}
\label{fig:adisr_com}
\end{figure*}

Figure \ref{fig:adisr_com} shows the disruption sizes of composite grains in which the maximum size is set to $100\mum$ and the mean wavelength of the radiation spectrum $\bar{\lambda}=1.2\mum$. In strong radiation fields ($U\sim 10^{3}-10^{8}$) near protostars, large aggregate grains can be rapidly disrupted into smaller grains with ice mantles. The range of the disruption size broadens when the tensile strength is lower because $a_{\rm disr}\propto S_{\max}^{-1/2}$ (see Eqs. \ref{eq:adisr_low} and \ref{eq:adisr_up}). One notes that in dense and cold regions (e.g., $n_{\H}\gtrsim 10^{9}\cm^{-3}$ and $T_{d}<100\K$), rotational disruption is inefficient for large grains ($a\gtrsim 1\mum$), such that grain coagulation to planetesimals is not affected by RATD. In this picture, we expect the maximum grain size is larger for dense conditions and weaker radiation fields.

\begin{figure*}
\centering
\includegraphics[scale=0.45]{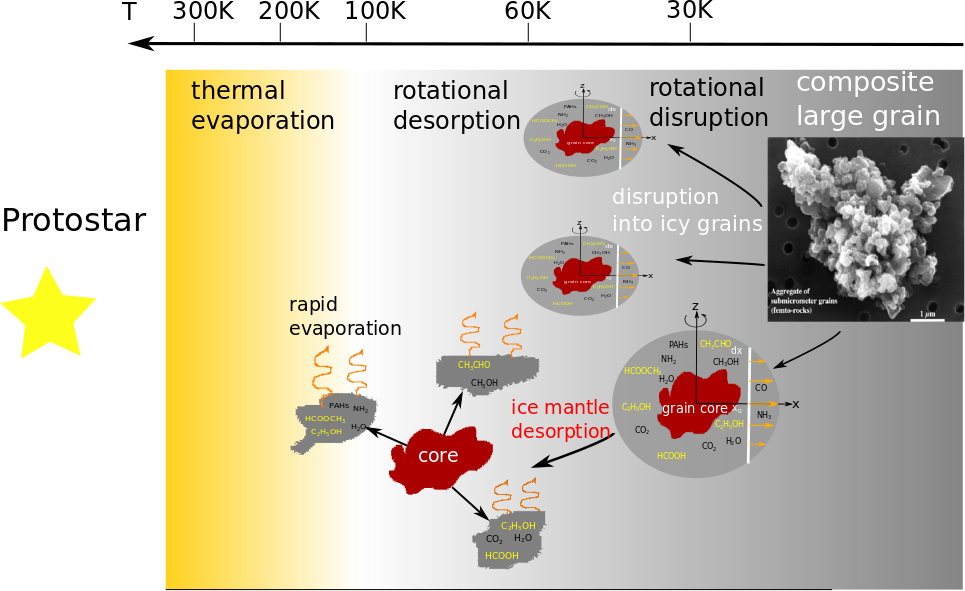}
\caption{A schematic illustration of dust evolution near a young stellar object: a big composite grain is first disrupted into icy individual grains and subsequent desorption of ice mantles into tiny fragments which are rapidly evaporated. Formation of COMs and water vapor is thus accompanied by destruction of large dust grains.}
\label{fig:hotcore}
\end{figure*}

Figure \ref{fig:hotcore} illustrates the evolution of composite grains near a young stellar object, starting with the disruption into small icy grains and subsequent desorption of the ice mantles into tiny ice fragments. The final stage involves rapid evaporation of mantle species into the gas phase due to higher temperatures of smaller grains by thermal spikes as well as the lower melting temperatures. 

In the rotational desorption paradigm, the release of COMs from ice mantles is accompanied with the variation of dust properties as a result of radiative torques. Thus, rotational desorption implies a correlation between complex molecules and dust properties (e.g., grain size distribution) because the intense radiation field that disrupts the ice mantles also disrupts large dust aggregates due to their low tensile strength.\footnote{The time difference between the disruption of aggregates and the desorption of ice mantles is rather short compared to the age of YSOs (years versus Myrs; see, e.g., Eq. \ref{eq:tdisr}). Therefore, in terms of observations, there would be no difference in the observing time of these effects.} This effect has a unique signature on observations. First, we expect the increased abundance of COMs corresponds to the reduction of large dust grains which implies lower opacity at mm-cm wavelengths. Second, the depletion of large aggregate grains results in the change in the polarization pattern because very large grains are expected to experience efficient self-scattering (\citealt{2015ApJ...809...78K}), whereas smaller ones are aligned along the radiation direction or magnetic field direction (\citealt{2007MNRAS.378..910L}; \citealt{2016ApJ...831..159H}; \citealt{2017ApJ...839...56T}; \citealt{Lazarian:2018vx}).

\subsection{Rotational desorption, COMs, snow-line, and dust properties in protoplanetary disks}
PPDs have layer structures where the gas density increases from $n_{\H}\sim 10^{5}\cm^{-3}$ in the disk atmosphere to $n_{\H}\sim 10^{12}\cm^{-3}$ in the disk plane. The gas corresponding temperature is $T\sim 1500\K$ to $T\sim 10\K$ in the disk plane (see e.g., Figure 2 in \citealt{Hoang:2018hc}). Therefore, we expect the rotational desorption to be efficient in the disk surface and intermediate layer where the density $n_{\H}<10^{10}\cm^{-3}$ and $T>100\K$. The proposed mechanism can explain the detection of COMs in PPDs (\citealt{2014A&A...563A..33W}; \citealt{Favre:2018kf}) as well as in protostellar disks \citep{2017ApJ...843...27L}. 

Furthermore, the sublimation of water ice occurs at $T\gtrsim 150\K$ which defines the snow-line. At this temperature, we show that ice mantles could already be disrupted into smaller fragments, resulting in subsequent evaporation into water vapor, provided that the local gas density is $n_{\H}\lesssim 10^{7}\cm^{-3}$ (see Figure \ref{fig:adisr_Td_coreshell_fixmantle}). As a result, water vapor can be observed in regions more extended than the traditional snow line of PPDs as constrained by water ice sublimation. As found in Tung \& Hoang (in preparation), the new snow-line desribed by rotational desorption is more extended than the classical one in the surface and intermediate layer of PPDs.

The present rotational mechanism can explain both the detection of COMs in low temperatures around PPDs and some correlation between dust polarization and COMs by ALMA \citep{Podio:2019dq}.
The authors found that the inner edge of H$_2$CO emission coincides with the location where the dust polarization pattern changes from parallel to the short axis to azimuthal direction, which is explained by polarization due to self-scattering by very large grains \citep{2018ApJ...865L..12B}. Moreover, the H$_2$CO abundance peaks at the edge of mm-dust emission, revealing that H$_2$CO may be increased due to the decrease of very large grains. Here, we suggest that large grains of $a\sim 10-100\mum$ are likely reduced from the location of H$_2$CO emission because of the decrease of gas density that facilitate RATD, such that self-scattering is reduced. But smaller grains can be radiatively aligned with the radiation direction (\citealt{2007MNRAS.378..910L}; \citealt{2016ApJ...831..159H}), producing an azimuthal polarization pattern (\citealt{2017ApJ...839...56T}). 

\section{Summary}\label{sec:summ}
We studied the effect of suprathermal rotation of grains on the desorption of ice mantles and COMs from grain surfaces in the environs of YSOs. Our principal results are summarized as follows.

\begin{enumerate}

\item
We showed that in the intense radiation of YSOs, the entire ice mantle can be disrupted into small fragments due to centrifugal force induced by extremely fast rotation of irregular grains spun-up by radiative torques. Thus, icy grain mantles would be removed before the grains can be heated to the ice sublimation temperatures of $T_{d}\gtrsim 100\K$.

\item We discussed the consequence of resulting fragments and find that the icy fragments of sizes $a\lesssim 10$\AAt~ are heated to high temperatures by single UV photons, triggering transient sublimation of COMs from the fragment. For larger fragments ($a>10$\AAt), thermal sublimation of COMs and water molecules is significantly enhanced compared to sublimation from the original grain because of the increase of grain temperature with decreasing grain size.

\item We identified some key difference between rotational desorption and classical thermal sublimation of COMs that can be tested with observations. We find that rotational desorption can be efficient at temperatures at least $\sim 20-40\K$ below the thermal sublimation threshold of COMs.

\item We applied the rotational desorption mechanism to study desorption of COMs in hot cores/corinos. We found that rotational desorption is efficient at much larger distances than thermal evaporation. This can successfully explain the observed COMs in cold, extended regions around high-mass/low-mass protostars.

\item We suggested that the observed enhancement of COMs in the outflow cavity walls could be induced by the rotational desorption mechanism due to irradiation of protostars through the outflow cavity and lower gas density compared to the disk regions. Rotational desorption is also important in the outflows of protostars and PDRs.

\item
We found that large aggregate grains which are presumably formed by coagulation in dense regions can be disrupted into distinct grains hosting ice mantles. Such icy grain mantles are then subsequently disrupted into smaller fragments, followed by rapid evaporation of COMs and water ice. Thus, we predict a correlation between COMs and the depletion of large grains in dense regions with strong radiation fields of YSOs.

\item
We discussed the implications of rotational desorption of ice mantles in PPDs. Due to rotational desorption, the snow-line of PPDs is expected to be more extended than determined by classical thermal sublimation. The observed correlation of COMs and dust polarization can also be explained by rotational desorption mechanism. 

\end{enumerate}

\acknowledgments

We are grateful to the anonymous referee for helpful comments that improved our manuscript. We thank Woojin Kwon for useful comments and suggestions. This work was supported by the National Research Foundation of Korea (NRF) grants funded by the Korea government (MSIT) through the Basic Science Research Program (2017R1D1A1B03035359) and Mid-career Research Program (2019R1A2C1087045). T.L.N is funded by the SOFIA Postdoctoral fellowship.

\bibliography{ms.bbl}

\end{document}